\documentclass[aps,prd,amsfonts,onecolumn,preprintnumbers,superscriptaddress,showpacs,floatfix,nofootinbib]{revtex4}

\usepackage{graphicx}

\newcommand{\be}{\begin{equation}} 
\newcommand{\ee}{\end{equation}}
\newcommand{\bea}{\begin{eqnarray}}
\newcommand{\eea}{\end{eqnarray}}

\newcommand{\gapp}{\mathrel{\raise.3ex\hbox{$>$}\mkern-14mu
              \lower0.6ex\hbox{$\sim$}}}
\newcommand{\lapp}{\mathrel{\raise.3ex\hbox{$<$}\mkern-14mu
              \lower0.6ex\hbox{$\sim$}}}

\newcommand\lsim{\lesssim}
\newcommand\gsim{\gtrsim}

\newcommand\vev[1]{{\langle {#1} \rangle}}
\renewcommand\({\left(}
\renewcommand\){\right)}
\renewcommand\[{\left[}
\renewcommand\]{\right]}

\newcommand\del{{\mbox {\boldmath $\nabla$}}}

\newcommand\eq[1]{Eq.~(\ref{#1})}
\newcommand\eqs[2]{Eqs.~(\ref{#1}) and (\ref{#2})}

\newcommand\eqsss[4]{Eqs.~(\ref{#1}), (\ref{#2}), (\ref{#3})
and (\ref{#4})}

\newcommand\eqst[2]{Eqs.~(\ref{#1})--(\ref{#2})}

\newcommand\eqreff[1]{(\ref{#1})}
\newcommand\eqsref[2]{(\ref{#1}) and (\ref{#2})}

\newcommand\pa{\partial}

\newcommand\mpl{M_{\rm P}}

\newcommand{\dlabel}[1]{\label{#1}}

\def\calb{{\cal B}}

\def\calp{{\cal P}}

\def\calv{{\cal V}}
\def\calpz{{\calp_\zeta}}

\newcommand\bfk{{\mathbf k}}

\newcommand\bfx{{\mathbf x}}
\newcommand\bfy{{\mathbf y}}

\newcommand\sunit{\,\mbox{s}}

\newcommand\TeV{\,\mbox{TeV}}

\newcommand\MeV{\,\mbox{MeV}}

\newcommand\msun{M_\odot}

\newcommand\sub[1]{_{\rm #1}}
\newcommand\su[1]{^{\rm #1}}

\newcommand\mone{^{-1}}
\newcommand\mtwo{^{-2}}
\newcommand\mthree{^{-3}}

\newcommand\half{^{1/2}}

\newcommand\threehalf{^{3/2}}

\newcommand{\fnl}{f\sub{NL}}

\newcommand\bfkp{{{\bfk}'}}

\newcommand\tpq{{(2\pi)^3}}

\newcommand{\calpzphi}{\calp_{\zeta_\phi}}
\newcommand{\calpzchi}{\calp_{\zeta_\chi}}
\newcommand{\calpzsigma}{\calp_{\zeta_\sigma}}
\newcommand{\phis}{\overline{\phi^2}}

\newcommand{\gsn}{g^2_{5}}
\newcommand{\zetaphi}{\zeta_\phi}
\newcommand{\zetasigma}{\zeta_\sigma}
\newcommand{\zetachi}{\zeta_\chi}
\newcommand{\phistar}{\phi_*}
\newcommand{\sigmastar}{\sigma_*}
\newcommand{\chistar}{\chi_*}
\newcommand{\rhophiend}{\rho_{\phi 2}}
\newcommand{\rhochiend}{\rho_{\chi 2}}
\newcommand{\phiff}{\phi_{55}}
\newcommand{\chiff}{\chi_{55}}
\newcommand{\Hff}{H_{55}}

\begin{document}

\pacs{98.80.Cq}

\title{
Preheating and the non-gaussianity of the curvature perturbation}
\author{Kazunori Kohri}
\affiliation{Department of Physics, Lancaster University, Lancaster LA1 4YB, UK}
\author{David H. Lyth}
\affiliation{Department of Physics, Lancaster University, Lancaster LA1 4YB, UK}
\author{Cesar A. Valenzuela-Toledo}
\affiliation{Escuela de F\'{\i}sica, Universidad Industrial de Santander,
Ciudad Universitaria, Bucaramanga, Colombia}

\begin{abstract}
The perturbation of a light field might affect preheating
and hence generate a contribution to the spectrum and non-gaussianity of
the curvature perturbation $\zeta$. 
The field might appear directly in the preheating
model (curvaton-type preheating) or indirectly through its effect on a
mass or coupling (modulated preheating).
We give  general expressions for $\zeta$ based on the $\delta N$ formula, and 
apply them to the cases of quadratic and quartic
chaotic inflation. For the quadratic case, curvaton-type preheating is 
ineffective in contributing to $\zeta$, but modulated preheating can 
be effective. For quartic inflation,  curvaton-type
preheating 
may be effective but the usual $\delta N$ formalism has to be 
modified. We see under what circumstances  the recent numerical simulation 
of Bond {\em et al.} [0903.3407] may be enough to provide a rough 
estimate for this case.\footnote
{This paper is dedicated to the memory of Lev Kofman who died on 12th November 2009}
\end{abstract}

\maketitle

\section{Introduction}

As far as we can tell, the observed cosmological perturbations originate
from a  primordial curvature perturbation $\zeta(\bfx)$, that is present
at the epoch when the shortest cosmological scale approaches horizon entry.
That epoch corresponds to a temperature of order $1\MeV$ and 
an age $t\sim 1\sunit$. At this stage the 
curvature perturbation is time-independent because the cosmic fluid is
radiation-dominated to very high accuracy, but it may be time dependent
at earlier times.

It is thought that $\zeta(\bfx,t)$  at each position is determined by
the values $\phi^*_i(\bfx)$,
of one or more scalar
fields   evaluated at some initial epoch during inflation.
(We shall consistently use a star to denote quantities evaluated at this
epoch.)
The corresponding perturbations $\delta\phi^*_i$
are   supposed to 
be generated from the vacuum fluctuation, which requires the fields to be
light.

According to the original paradigm, $\zeta$ is generated entirely
from the perturbation $\delta\phi_*$
of the inflaton field  in a slow roll inflation
model. In that case, $\zeta$ is present already at the initial epoch, 
being constant thereafter.
The  non-gaussianity of $\zeta$
is in this case very small, and almost certainly undetectable.

According to  alternative paradigms, the perturbation $\delta\chi_*$ of
a light field other than the inflaton  generates  a significant
(maybe dominant)  contribution.
 Such a contribution is initially negligible, growing to its final
value later,  and in general only after inflation is over. The non-gaussianity in this
case can be detectable.

In this paper we consider the growth of $\zeta$ that may occur  during 
preheating. 
Preheating may be defined as the loss of energy by time-dependent scalar
fields, through mechanisms other than 
single particle decay. 
Preheating is typically followed by reheating,
with an intervening era between the two. Reheating may be defined as the
practically complete
thermalization of a gaseous cosmic fluid, including the case that the
constituents of the gas correspond to an oscillating scalar field.

The simplest mechanism for preheating is the potential
\be
V =  \frac12 m^2 \phi^2  + \frac12 g^2 \phi^2 \chi^2 
\dlabel{pot2},
\ee
with $\phi$ the inflaton field, and $\chi=0$ during inflation
so that we deal with the quadratic (`chaotic') inflation potential.
If no other terms of the action are relevant, $\phi$ oscillates
and loses energy 
by creating $\chi$ particles through what is called parametric
resonance  \cite{bt,kls,bt2,kls2}. As in this example, the field $\chi$
is usually different from the field $\phi$, but parametric resonance works in just
the same way if they are the same. It also works with different forms of the 
potential, such as
\cite{kls3}
\be
V =  \frac14 \lambda \phi^4  + \frac12 g^2 \phi^2 \chi^2 
\dlabel{pot3} ,
\ee
with again $\phi$ the inflaton field. (That case is often called massless
preheating.) Given a potential that allows preheating, it can happen that
the produced field $\chi$  can decay sufficiently rapidly into (say) a pair
of fermions, so that $\phi$ can lose its energy before even one oscillation 
takes place. That is called  instant preheating \cite{fkl}.  
There is also what is known as tachyonic preheating \cite{tachyonic},
which in its simplest form invokes  the potential
\be
V(\chi) = V_0 - \frac12m^2\chi^2 + \frac14\lambda\chi^4
. \dlabel{pot4} \ee 
Tachyonic  preheating occurs because the 
 effective mass-squared $-m^2$ is initially to be positive. During the
sign change of $m^2$  the vacuum 
fluctuation of $\chi$ to be promoted to a classical perturbation which is then
amplified as $\chi$ rolls off the hilltop. Parametric resonance can take place
is $\chi$ passes through the minimum, followed by  
more amplification of the original perturbation and so on.

These  basic  preheating mechanisms   have been considered within
several different scenarios. In most of them,
 preheating takes place immediately  after inflation 
so that the oscillating field  is the one 
 involved in the inflation model. (If the inflation model is hybrid or multi-field
there are two or more of these fields.)
In some  scenarios
 though,  preheating takes place after a later phase transition and  the 
 oscillating field played   no role during inflation.

As we will explain, any growth of $\zeta$ occurring during preheating may
persist for some time afterwards. The growth will generally
terminate at or before reheating. (If there is further growth
during   or after
reheating then that growth should be treated as a separate process.)
For clarity we will just talk about `growth during preheating' on the 
understanding that any subsequent growth prior to reheating is to be included
in the discussion.

Regarding its  ability to  generate a contribution to $\zeta$,
preheating has some points of similarity with reheating. Two distinct
mechanisms exist in the reheating case. In the original `curvaton'
mechanism\footnote
{The mechanism was also  proposed in the context of a bouncing  universe;
see \cite{es} for a discussion and earlier references.} 
 \cite{curvaton1,curvaton2,curvaton3,mollerach,luw},
the light field responsible for generating the contribution to $\zeta$ 
is the oscillating scalar field, whose decay is responsible for reheating.
Later, the `modulated reheating' mechanism was proposed
\cite{inhomreh,zaldarriaga}.  There,
the relevant light field acts only indirectly, by  affecting the decay
rate of the inflaton
 so that we have $\Gamma(\chi(\bfx))$ where that $\Gamma$  is the
decay rate and $\chi$ is the perturbed light field.
With this in mind,  we can distinguish two versions of the preheating
scenario. In a `curvaton-type preheating' scenario
the light field is directly involved in  preheating; either a field  created by preheating,
or an oscillating field that is responsible for the creation.
In a `modulated preheating' scenario the light field instead acts
indirectly, by affecting a preheating parameter such as
the coupling $g$ in \eq{pot2}. In that case we have $g(\chi)$
where $\chi$ is the light field. 

Modulated preheating or reheating can be implemented within any 
scenario, but becomes predictive only when one specifies the dependence
of the parameter on the field (ie.\ the function $\Gamma(\chi)$
or $g(\chi)$). 
Curvaton reheating and curvaton-type preheating are more restricted,
but  more predictive because $\Gamma$, $g$ and so on are taken to be
constants. We shall show that curvaton-type preheating cannot occur
with the potential \eqreff{pot2} if it is supposed to hold also during
inflation.

What we are calling modulated preheating has not so far been studied, though
its possibility has been recognized \cite{acker}.  (For related works 
see \cite{batt}.) 
The possibility of 
what we are calling curvaton-type preheating has been considered in a few
papers.  The papers \cite{anupamkari1,anupamkari2}, 
consider the  quadratic inflaton potential \eqreff{pot2}, 
 taking the potential  to be valid also during inflation.
The  papers
 \cite{anupamkari2,anupamkari3,barnabycline}  consider  tachyonic preheating. 
The papers   \cite{instant} consider  a multi-field generalization
of \eq{pot2} with instant preheating. 
The papers  \cite{massless1,anupamasko,massless2,lev} consider  the 
massless preheating
scenario of \eq{pot3}. Reference \cite{bdk} considers preheating
at the end of a hybrid inflation model involving the Higgs field. 
In these works, preheating starts as soon as inflation is over.
In another work \cite{cnr}  the  curvaton field causes preheating, when it starts
to oscillate long after inflation is over. In 
 \cite{johntoni} preheating is caused by the oscillation of a
  flat direction of the MSSM, again long after inflation is over.

Most of the  papers cited above 
 use cosmological perturbation theory, at either first
 order \cite{instant,barnabycline,johntoni,bdk} or second order
\cite{anupamkari1,anupamkari2,anupamkari3,anupamasko}.
We  instead use the $\delta N$ formalism  \cite{ss,lms,lr}.
We  begin in Section \ref{scp} by recalling the 
basic properties of the curvature perturbation.
In Section \ref{sdn} we carefully set up  the $\delta N$ formalism.
In Section \ref{schilight}    we show that
the quadratic potential \eq{pot2} cannot provide  
curvaton-type preheating, contrary to what was assumed in \cite{anupamkari1,anupamkari2}. 
In Section \ref{sec:sip} we consider modulated
preheating with the quadratic potential.
In Section \ref{sctp} we consider curvaton-type preheating 
the quartic potential \eqreff{pot3} (massless preheating).
In contrast with other applications of the $\delta N$ formalism,
the perturbations of the light fields generated after cosmological scales leave the horizon
are likely to  be significant in this case. Ignoring them, we show how to use 
the recent numerical simulation of \cite{lev} to  give an order of magnitude estimate of
$\fnl$ with the unperturbed value of $\chi_*$ set equal to zero. It disagrees with the estimate
obtained in \cite{anupamasko} using second order cosmological perturbation theory.
 In a concluding section we summarise our
finding, and point to future directions for research. In two appendices
we extend the discussion of Section \ref{schilight}.

\section{The curvature perturbation} 

\dlabel{scp}

\subsection{Definition and evolution}

{} The non-perturbative definition of 
the primordial curvature perturbation $\zeta$ is described  in for instance 
\cite{newbook}, where original references can be found.
The components of the metric tensor are smoothed on  a comoving scale $R$
and one considers  the super-horizon 
regime $aR> H\mone$ where $H\equiv \dot a/a$ is the Hubble parameter
and $a(t)$ is the scale factor normalised to 1 at present.\footnote
{Smoothing a function $g(\bfx)$ means that
$g$ at each location is replaced by its average within a sphere of coordinate
radius $R$ around that position. The averaging may be  done with a smooth
window function such as a gaussian. The smoothed function is supposed to have 
no significant Fourier components with coordinate wavenumbers  $k\gg R$, which means that its gradient
at a typical location is at most of order $1/R$. A function $g$ with that property
is said to be `smooth on the scale $R$'.} The energy density $\rho$ and pressure $P$
are smoothed on the same scale.
One  considers the  slicing of spacetime with uniform energy density.
 The spatial metric  is  written as
\be
g_{ij}(\bfx,t) \equiv a^2(t)e^{2\zeta(\bfx,t) } \(I e^h(\bfx,t) \)_{ij}
, \dlabel{gij} \ee
where  $h$ is traceless so that $e^h$ has unit determinant. 
The smoothing scale is chosen to be somewhat shorter
than the scales of interest, so that the Fourier components of $\zeta$ on those scales
is unaffected by the smoothing.  The threading of spacetime is taken to be orthogonal
to the slicing. The
time dependence of the locally defined scale factor
$a(\bfx,t)\equiv a(t) \exp(\zeta)$ defines the rate at which an infinitesimal
comoving  volume $\calv$ expands:
$\dot \calv/\calv=3\dot a(\bfx,t)/a(\bfx,t)$.

Under the reasonable assumption that the Hubble scale $H\mone$ is the biggest relevant
distance scale, 
the  energy continuity equation $d(\calv \rho)= - P d\calv$ at each 
location is the same as in a homogeneous universe; as far as the evolution of $\rho$
is concerned, we are dealing with a family of separate homogeneous universes.
 With the additional assumption that the initial condition is set
by scalar fields during inflation (adopted here),  the smoothed 
$h_{ij}(t)$ is time-independent  after 
smoothing and then the separate universes are  homogeneous as well as isotropic.

Since we are working on slices of uniform $\rho$, the energy continuity equation can
be written 
\be
\dot \rho(t) = - 3 \[ H(t)  + \dot \zeta(\bfx,t) \]
\[ \rho(t) + P(\bfx,t) \]
. \dlabel{econt} \ee
One write
\be 
P(\bfx,t) = P(t) + \delta P(\bfx,t) 
, \ee
so that $\delta P$ is the pressure perturbation on the uniform density slices, 
and choose $P(t)$ so that the unperturbed quantities satisfy the unperturbed  equation
$\dot\rho = -3H(\rho + P)$. Then
\be
\dot \zeta = - \frac{H\delta P}{\rho + P + \delta P}
. \dlabel{zetadot} \ee
This gives $\dot\zeta$ if we know $\rho(t)$ and $P(\bfx,t)$. It makes
 $\zeta$  time-independent  during any  era when  $P$
 is a unique function of  $\rho$ 
\cite{earlier,lms} (hence uniform
on slices of uniform $\rho$).  The pressure 
perturbation is  said to be adiabatic in this case, otherwise it is said to be
 non-adiabatic.

The key assumption in the above discussion is that 
in the superhorizon regime certain smoothed quantities
(in this case $\rho$ and $P$)   evolve at each location as they would in an unperturbed
universe. 
In other words, the evolution of the perturbed universe is that of a family of unperturbed
universes. This is the separate universe assumption, that is useful also in other situations
\cite{newbook}. 

The primordial curvature perturbation $\zeta$
is directly probed by observation on `cosmological
scales' corresponding to  roughly 
$e^{-15} H_0\mone \lsim k\mone \lsim H_0\mone$.
These scale begin to enter the horizon when
 when $T\sim 1\MeV$. The  Universe at that stage
is  radiation dominated to very high accuracy, 
implying $P=\rho/3$ and a constant curvature perturbation
which we denote simply by $\zeta(\bfx)$.
When cosmological scales are the only ones of interest, one should
choose the smoothing scale as $R\sim e^{-15} H_0\mone$. Unless stated otherwise, we make
this choice.

Within a given scenario,
$\zeta$ will exist also on smaller inverse wavenumbers, down to some `coherence length'
which might be as low as $k\mone \sim e^{-60}H_0\mone$ (the scale leaving the horizon at the
end of inflation). If one is interested in such scales, the smoothing scale $R$ should be
chosen to be (somewhat less than) the coherence length. 

\subsection{Correlators}

Theories will predict the correlators of  cosmological perturbations.
The two-point correlator of a  perturbation $g(\bfx)$
is written $\vev{g(\bfx_1) g(\bfx_2)}$ and so on for higher correlators.
The bracket basically denotes  an ensemble average,
with our universe a typical member of the ensemble.
If the perturbations originate as a vacuum fluctuation during inflation,
the ensemble average is the vacuum expectation value.
If the universe is homogeneous this makes the correlators
 translation invariant, which we will assume.\footnote 
{The quantity  $\vev{g(\bfx)}$ is then  just 
 a number which can be set to zero by choice of the unperturbed quantity.}

 Given translation invariance,
 the ergodic theorem holds whereby the bracket can be taken to be 
a spatial average. Thus
 $\vev{g(\bfx_1) g(\bfx_2)}$ is the spatial average of 
$g(\bfx_1)g(\bfx_2)$ at  fixed $\bfx_1-\bfx_2$, 
within any  volume with size $L$ much bigger than $|\bfx_1-\bfx_2|$.
For any finite $L$, the volume average differs from the the ensemble average.
The difference is called cosmic variance.

To exploit the translation invariance it is convenient to work with
Fourier components $g_\bfk$. Using a box of coordinate size $L$ we have 
\bea
g_\bfk &=& \int d^3x g(\bfx) e^{-i\bfk\cdot\bfx} 
. \dlabel{fourdef} \\
g(\bfx) &=& L\mthree \sum_\bfk g_\bfk e^{i\bfk\cdot\bfx} 
\simeq (2\pi)\mthree \int d^3k g_\bfk e^{i\bfk\cdot\bfx} 
. \eea
Since the box imposes a periodic boundary condition, only the wavenumbers 
$kL \gg 1$ are physically significant. For them one can use the 
final equality.  According to the  ergodic theorem, 
 one can regard $\vev{g_{\bfk_1}g_{\bfk_2}}$ as an average over the points
within a cell of  momentum space, up to the cosmic variance.

The dependence of (the loop contributions to) the correlators
upon the box size, reflects the fact that the correlators are in position space are spatial 
averages within the box.
Observations are available  within
the observable universe, with coordinate size of order 
 Except for the low multipoles of the CMB, all observations
probe scales $k\gg H_0$. To handle them, one should choose the box size as
$L=H_0\mone$ \cite{leblond}. A 
 smaller choice would throw away some of the data while
a bigger choice would make the spatial averages unobservable.

Low multipoles of the CMB anisotropy 
 explore scales of order $H_0\mone/\ell$ not very much smaller than
$H_0\mone$. To handle them  one has to take $L$ bigger than $H_0\mone$.
For most purposes, one should use a  box, such that $\ln(LH_0)$ is just
a few (ie.\ not exponentially large)
 \cite{mybox,davids,newbook}. 

To summarise, the box size should normally be taken as $L=H_0\mone$. The exception
is the case of low CMB multipoles, where one should normally take $L\gg H_0\mone$
with $\ln(LH_0)$ not exponentially large. A box satisfying these requirements will
be called a minimal box.

The spectrum $P_g$ is defined in terms of the two-point correlator by
\be
\vev{g_\bfk g_\bfkp} =
(2\pi)^3 \delta^3(\bfk+\bfk') P_g(k)  \dlabel{fourspec}
. \ee
One also defines $\calp_g(k) \equiv (k^{3}/2\pi^{2})P_g(k)$,
also called the spectrum. 
An equivalent expression \cite{newbook} is
\be
\vev{g(\bfy) g(\bfx+\bfy)} = (2\pi)\mthree \int
d^3k P_g(k) e^{i\bfk\cdot\bfx}
, \dlabel{spatialaverage} \ee
where the bracket can be regarded as 
the spatial average with respect to $\bfy$.

The mean-square perturbation, evaluated within a  box of size $L$
and smoothed on scale $R$,   is
\be
\vev{g^2(\bfx)} = \int^{R\mone}_{L\mone} \frac{dk}k \calp_g(k)
\simeq \calp_g \ln(L/R)
. \dlabel{meansquare}
\ee
{} The final estimate is valid if $\calp_g(k)$ is more or less 
scale-independent as will usually be the case for primordial perturbations.
The spectrum of $\pa g/\pa x_i$ (the derivative with respect to one of the coordinates)
is $k_i^2\calp_g$, hence the mean-square gradient is
\be
\vev{ (|\del g|)^2 } = \int^{R\mone}_{L\mone} \frac{dk}k k^2 \calp_g(k)
\simeq \frac12  \calp_g/R^2 \sim \vev{g^2}/R^2
. \dlabel{msgradient} \ee
{} 

For a gaussian perturbation, the spectrum defines all correlators.
The $n$-point correlators  vanish  for odd $n$, while for even $n$
\be
\vev{g_{\bfk_1} g_{\bfk_2} g_{\bfk_3} g_{\bfk_4}} =
\vev{g_{\bfk_1}g_{\bfk_2}}\vev{g_{\bfk_3} g_{\bfk_4}}
+\vev{g_{\bfk_1}g_{\bfk_3}}\vev{g_{\bfk_3} g_{\bfk_2}}
+\vev{g_{\bfk_1}g_{\bfk_4}}\vev{g_{\bfk_2} g_{\bfk_3}}
, \dlabel{fourpoint} \ee
and so on. For a non-gaussian contribution one has to specify more quantities,
starting with the  bispectrum $B_g$:
\bea
 \langle g_{\bfk_1}  g_{\bfk_2}  g_{\bfk_3} \rangle
&=& \tpq \delta^3(\bfk_1+\bfk_2+\bfk_3) B_g(k_1,k_2,k_3) \dlabel{bispec} \\
 B_g(k_1,k_2,k_3) &=&\calb_g(k_1,k_2,k_3) \[ P_g(k_1) P_g(k_2) +\,{\rm
cyclic\ permutations} \]  . \dlabel{fnldef} \eea
The quantity  $\calb_g$ is called the reduced bispectrum.
For $\zeta$ one uses  $\fnl\equiv (5/6)\calb_\zeta$.

{}From observation of the CMB anisotropy and the galaxy distribution,
we know that  $\calp_\zeta(k)$  is 
almost scale-independent
with the  value $(5\times 10^{-5})^2\sim 10^{-9}$.   {}
Also, we know that $\zeta$ is gaussian
to high accuracy. Taking 
$\fnl$ to be practically scale
independent (`local' form)  as is predicted by the models that we consider, the
bound at $95\%$ confidence level according to the first of \cite{curto}  is
$-4 < \fnl < 80$. 


\section{The $\delta N$ formula}

\dlabel{sdn}

\subsection{The general formula}

\dlabel{sgeneralformula}

Consider now,  the analogue of \eq{gij} for a generic slicing:
\be
g_{ij}(\bfx,t) \equiv a^2(t)e^{\psi(\bfx,t) } \tilde \gamma_{ij}(\bfx,t)
, \dlabel{gij2} \ee
where $\tilde \gamma$ has unit determinant. Take the initial
slice to be flat (meaning that  $\psi=0$) and the final slice to be
of uniform energy  density so that $\psi=\zeta$. This gives
 \cite{lms}
\be
\zeta(\bfx,t)=\delta N(\bfx,t)
, \dlabel{deltanformula} \ee
 where $N $ is the number of 
$e$-folds of expansion between the initial and final slice.
It is independent of the choice of the initial flat slice
because the expansion going from one flat slice to another is 
uniform. 
By virtue of the separate universe assumption, this formula allows
one to calculate $\zeta(\bfx,t)$ given the evolution of the scale factor in some family of 
unperturbed universes. 

To proceed, we invoke light fields, taken to be canonically normalized.
Ignoring mixed derivatives for simplicity, a  light field $\phi_i$ is defined
\cite{newbook} as one  whose
 effective mass-squared  $m_i^2$ during inflation satisfies 
\be
|m_i^2| \ll H^2,\qquad m_i^2 \equiv \pa^2 V/\pa \phi_i^2
. \dlabel{lightcon} \ee
As each scale $k$ leaves the horizon during inflation, the  vacuum fluctuation
of each light field is converted to a classical perturbation \cite{newbook,ls}
. At a given epoch during inflation, 
the classical fields $\phi_i(\bfx)$ are  therefore smooth on the horizon scale. 

Now we take the initial epoch to be the one when the smoothing 
 scale $R$ leaves the horizon, and 
make a  crucial assumption.  {\em At 
 each location, one or more of the   light fields 
  provides  the initial
condition for the evolution of the local scale factor}
 (along with unperturbed quantities 
including parameters of the field theory and the values of any relevant unperturbed
fields).  Then we have {}  \cite{lr}
\be
\zeta(\bfx,t) =\delta N
\equiv N(\rho(t),\phi_1^*(\bfx),\phi_2^*(\bfx),\cdots) - 
N(\rho(t),\phi_1^*,\phi_2^*,\cdots) 
, \dlabel{basicdeltan} \ee
where the  star denotes the field values at the initial epoch. 
 Since the typical magnitude of a light
field perturbation is of order $H_*$, we need 
values of $\phi_i(\bfx)$ 
in  a  range  $\Delta\phi_*\sim  (H_*/2\pi)$,  centred on the unperturbed
values  $\phi_*$. 

These  initial values  $\phi_i^*(\bfx)$ 
are smooth on the scale $R$, and so are the initial values
of $\rho$ and $P$. If the subseqent evolution of the light fields were classical,
they would remain smooth on the scale $R$, and so would
$\rho$ and $P$. In fact, the 
 vacuum fluctuation  continues to generate  classical field perturbations, so that by
the end of inflation they are present on scales  $(aH)\sub{end}\mone < k\mone <R $.
In all known cases except that of massless preheating, 
these smaller scale perturbations in the light field have little  effect on 
the perturbations in $\rho$ and $P$ on 
scales of interest $k\mone> R$ and so can be ignored.
 Then, by virtue of the separate universe
assumptions,  
the evolution of $\rho$ and $P$ at each location is the same as in an unperturbed
universe, and so is the evolution of $N$. We can evaluate the curvature perturbation
by considering a family of unperturbed universes!
 This is the usual version of the $\delta N$ formalism,
 which we use except
when dealing with massless preheating.  

\subsection{Power series in the field perturbations}

For each field we can write
\be
\phi^*_i(\bfx) = \phi^*_i + \delta\phi^*_i(\bfx)
.  \ee

In all cases so far considered,
 we can expand   $N(\rho(t),\phi_1^*(\bfx),\phi_2^*(\bfx),\cdots)$ 
as a low-order power series  about the unperturbed field values:
\be
\dlabel{eq:zeta2nd}
\zeta(\bfx,t) = \sum_i N_i(t) \delta\phi^*_i(\bfx) + 
\frac12 \sum_{i,j} N_{ij}(t) 
\delta\phi^*_i(\bfx)\delta\phi^*_j(\bfx) + \ldots, 
\ee
where a subscript $i$ 
denotes $\pa /\pa \phi^*_i$ evaluated at the unperturbed
point of field space. To get the observed quantity $\zeta(\bfx)$ this should be 
evaluated after the $N_i(t)$ and so on settle down to their final values,
which we denote simply by $N_i$ and so on.

This formula  was first given in \cite{lr} following the non-linear proof in
\cite{lms} of $\zeta=\delta N$. 
To first order in the field perturbations, the formula and the proof
were  given in \cite{ss}, which focuses on the case that 
$\zeta$ is generated during inflation.
Related  formulas for the evolution of $\zeta$ during inflation 
were given in \cite{starobinsky} (first  order)
and in \cite{sb} (non-linear). 

The  light field perturbations  have completely negligible non-gaussianity
\cite{mald},
and their  spectrum soon  after horizon exit for the scale $k$ is given by
\cite{newbook}
\be
\calp_{\delta\phi_i}(k) \simeq (H_k/2\pi)^2
, \dlabel{calpdeltaphi} \ee
where $H_k$ can be taken as the value of $H$ at horizon exit.
The  observed near-gaussianity of $\zeta$ requires that \eq{eq:zeta2nd}  
is well approximated by one or more linear terms, with the field perturbations
nearly gaussian.  Taking the initial epoch to be just after horizon exit
we then have
\be
\calpz(k) \simeq \sum N^2_i  \( \frac{H_k}{2\pi} \)^2
. \dlabel{genspec} \ee

Assuming that the reduced bispectrum $\fnl$ 
is big enough to be observable (say $|\fnl|\gsim 1$), it is generated by
 the quadratic terms, the slight non-gaussianity of the $\delta\phi_i$ being 
negligible.
In this case, $\fnl$  is almost  scale-independent (local form). 

\subsection{The inflaton contribution to $\zeta$}

\dlabel{stwofield}

We will suppose that  there is  standard
slow-roll  inflation, which is effectively single-field because
only the inflaton field $\phi$ has significant variation during inflation.

The slow-roll approximation gives
 $V\simeq 3\mpl^2H^2$ and $\dot\phi\simeq -V'/3H$, where $V(\phi)$ is the 
potential during slow-roll inflation.  This gives the well-known relation
\be
N(k) = \mpl\mtwo \int^{\phi(k)}_{\phi\sub{end}} \frac V{V'} d\phi
, \dlabel{usualN} \ee
where  $N(k)$ denotes the number of $e$-folds of  slow-roll inflation 
occurring after the epoch 
of horizon exit for the  scale $k\mone$.
The inflationary potential will determine the field  value $\phi\sub{end}$ at which slow-roll
inflation ends.

One usually denotes $N(H_0)$ simply by $N$. It  is  determined by the inflationary
energy scale $H_*$ and the  post-inflationary evolution of the scale factor.
A typical value, which we adopt, is  $N\simeq 55$. For a  smaller scale, $N(k)
=N-\ln(k/H_0)$. For the  smallest cosmological scale $k\sub{min}\mone$
(enclosing mass $10^5\msun$ or so) $N(k) = N-15=40$. The initial epoch, when
the value of $\phi$ is denoted by $\phi_*$,  can be soon after that.
  The important feature of \eq{usualN} is
that {\em it determines the unperturbed field value $\phi_*$ and hence $\calpzphi$,
given  the inflationary potential and the 
post-inflationary cosmology}. 

Perturbing  $\phi$ with any other light fields fixed gives a constant contribution 
$\zeta_\phi(\bfx)$ to
$\zeta$, because it represents a shift back and forth along the
trajectory which does not alter the subsequent relation between $\rho$
and $P$.  The  perturbation $\delta\phi_*$ has typical
magnitude $H_*/2\pi \ll \phi_*$.
Since $\zetaphi$ represents a shift along the inflaton trajectory,
$N_\phi = - H/\dot\phi$ evaluated at the initial epoch.  Using the slow-roll
approximation one finds that $\zetaphi$ is practically linear, making it practically
gaussian \cite{slfield}. Taking the initial epoch soon after horizon exit,
\be
\calpzphi(k) = N_\phi^2 (H_k/2\pi)^2
. \label{calpzphi}
\ee
If $\calpzphi$ dominates, the spectral tilt  is $n-1=2\eta-6\epsilon$,
where $\eta\equiv \mpl^2 V''/V$ and $\epsilon\equiv \mpl^2(V'/V)^2/2$
evaluated at horizon exit.

\subsection{The non-inflaton  contribution}

\label{snoninf}

Now we consider the contribution of light fields other than the inflaton.
We suppose that only one of them contributes significantly, and call it $\sigma$.
 Its  contribution  $\zeta_\sigma(\bfx,t)$ is 
initially negligible, but it can grow by generating a non-adiabatic
pressure perturbation. This might happen during any era, except one of
complete matter domination ($P=0$) or complete radiation domination
($P=\rho/3$).  Possibilities in chronological order include generation
 (i) during multi-field inflation  \cite{starobinsky},  (ii) at the end
of inflation \cite{myend}, (iii) during preheating as we discussed in the
Introduction, (iv) during  
modulated reheating  \cite{inhomreh,zaldarriaga,acker,Suyama:2007}, (v) at a 
modulated phase transition~\cite{Matsuda:2009yt}  and (vi) before
a  second reheating through the curvaton mechanism \cite{luw}. In this
paper we focus on preheating.

We interested in the value of $\zetasigma$ after it   settles down to its final constant
value, which we  denote  simply by $\zetasigma(\bfx)$
\bea
\zetasigma(\bfx) 
&\equiv&  \zeta(\bfx) -\zeta_\phi(\bfx) \\
&=& \delta N(\sigmastar(\bfx),\phi_*)
+ \delta 
\[ N(\sigmastar(\bfx),\phistar(\bfx))) - N(\sigmastar(\bfx),\phistar) - N(\sigmastar,\phistar(\bfx)) \]
. \dlabel{zetasigma4} \eea
We make the followign approximation, which is usually adequate:
\bea
\zetasigma(\bfx) &\simeq& \delta N(\sigmastar(\bfx),\phi_*) \dlabel{zetasigma3} \\
&\simeq&   N_\sigma \delta\sigmastar(\bfx) + \frac12 N_{\sigma\sigma} \( \delta\sigmastar(\bfx) \)^2
. \dlabel{zetasigma2}
\eea
\eq{zetasigma3} ignores mixed derivatives  $\pa^2 N/\pa \phistar\pa \sigmastar$ 
and so on, while \eq{zetasigma2} ignores  $\pa^3 N/\pa \sigmastar^3$ and higher derivatives.

The   unperturbed value  $\sigmastar$ is  
defined as the   spatial average of $\sigma(\bfx)$ within the
minimal box. It  is   in general  a free
parameter whose value has  to be specified along with the
parameters of the Lagrangian.
To get some idea about is  likely value though,
we can use the fact that the mean square perturbation, 
smoothed on scale $L_1$ and evaluated within a larger box of size
$L_2$, is   
\be
\overline { (\delta\sigmastar)^2 } = \int^{L_1\mone}_{L_2\mone} \frac{dk}k 
\calp_{\delta\sigmastar} (k) \sim (H_*/2\pi)^2 \ln(L_2/L_1) 
, \dlabel{superlarge} \ee
where in the final equality we set $\calp_{\sigmastar}\sim (H_*/2\pi)^2$. 
 To use this formula we can take 
 $L_1=L$ with $L$ a minimal box centred on the observable universe.  
 Then the formula says that the unperturbed value of $\sigmastar$ 
(ie.\ it's spatial average within the minimal box) will 
be at least  of order $H_*/2\pi$ if the observable universe is located
at a typical position. If there were a very large number
of $e$-folds of almost-exponential 
inflation before the observable universe left the horizon.
we can go further; taking the formula literally
we can choose  $\ln(L_2/L_1)$ to be much bigger than 1,
 and  conclude that the  value of the unperturbed field
for a typical location of our universe will be much bigger than
 $H_*/2\pi$. (As $\ln(L_2/L_1)$ is increased the calculation leading
to \eq{superlarge} will at some stage become out of control
\cite{mybox} but we don't need an enormous value of $\ln(L_2/L_1)$
to arrive at the above conclusion.) Of course, it might be that we live in 
a special place where the the unperturbed $\chi_*$  is much less than
$H_*/2\pi$.

Using \eqsss{fourdef}{fourspec}{fourpoint}{fnldef},
 and taking $\delta \sigma_*$ to be gaussian with a scale-independent  
spectrum 
$\calp_{\delta \sigma_*} = (H_*/2\pi)^2$,  we can evaluate the
spectrum and bispectrum of the curvature perturbation.  One finds
\bea
\calpzsigma &=&\calpzsigma\su{tree} +  \calpzsigma\su{loop} 
\dlabel{first1} \\
\calpzsigma\su{tree} &=& N_\sigma^2 \calp_{\delta\sigma_*}
= N_\sigma^2 (H_*/2\pi)^2 \dlabel{treespec} \\
\calpzsigma\su{loop}(k) &=& 
\frac14 N^2_{\sigma\sigma}  \calp_{(\delta\sigma_*)^2}(k) 
= N^2_{\sigma\sigma}  \ln(kL) \( \frac{H_*}{2\pi} \)^4
, \dlabel{ploop} \eea
and 
\bea
\fnl &=& \fnl\su{tree} + \fnl\su{loop} \\
\fnl\su{tree} &=& \frac56 \( \frac{\calpzsigma}{\calpz} \)^2 
\frac{N_{\sigma\sigma}}{N_\sigma^2} \dlabel{fnltree} \\
\fnl\su{loop}(k_1,k_2,k_3) &\sim& 
\ln(kL) \frac{ N_{\sigma\sigma}^3}{\calpz^2} \( \frac{H_*}{2\pi} \)^6
 \dlabel{last}
.\eea

The labels `tree' and `loop' refer to the Feynman-like diagrams of
\cite{bksw}. 
The loop contributions correspond to formally divergent
integrals, but they are regularized by setting $\calpzsigma=0$ 
for $k\lsim L\mone$, with $L$ the  box size.
\eq{ploop}  was given in \cite{myaxion}.
It is valid  for $k\gg L\mone$, which is anyhow 
required so that one can ignore the artificial periodicity of the Fourier
series.  \eq{fnltree} was first given in \cite{fnllocal}.
\eq{last}  was given in
\cite{bl}, with $k_i$ taken to have   a common value $k\gg L\mone$
to rough order of magnitude. 

We see that
\bea
\calpzsigma\su{loop} &=&
\ln(kL) \( \frac {H_*}{2\pi} \frac{N_{\sigma\sigma}}{N_\sigma} \)^2
\calpzsigma\su{tree} , \\
\fnl\su{loop} &\sim & \ln(kL) \( \frac {H_*}{2\pi} \frac{N_{\sigma\sigma}}{N_\sigma} \)^2
\fnl\su{tree} \dlabel{fnlloop2}
. \eea
By examining the integral \cite{bl} leading to \eq{last}, one can see that 
\eq{fnlloop2} 
becomes exact in the squeezed configuration  $k\equiv k_1\ll k_2\simeq k_3$. 
This has  not been noticed before, and is quite interesting because
the observational constraint on a local $\fnl$ comes mostly from the 
 squeezed configuration  \cite{curto}.

We see that the tree contributions dominate if
\be
\ln(kL) \(  \frac{H_*}{2\pi} \frac{ N_{\sigma\sigma} }{N_\sigma } \)^2
\ll  1, \ee
while the loop contributions dominate in the opposite case.
An equivalent statement is that the tree contributions dominate
if the linear term of \eq{zetasigma2} dominates, while the loop
contributions dominate if the quadratic term of \eq{zetasigma2} dominates.
(The statements are equivalent because the mean-square
of   $\delta \sigma_*$, smoothed on the scale $k$, is
$\ln(kL) (H_*/2\pi)^2$.)

When comparing the loop contribution with observation one should normally
set $L=H_0\mone$,
except for the low CMB multipoles where one should choose $L\gg H_0\mone$ with 
$\ln(kL)\sim 1$. With the choice $L=H_0\mone$, $\ln(kL)\sim 5$ for the scales
explored by the CMB multipoles with $\ell \sim 100$, while $\ln(kL)\sim 10$
for the scales explored by galaxy surveys. This increase would be observable
\cite{leblond}.

If $\zetasigma\simeq \zeta$ dominates, the near-gaussianity of $\zeta$ requires \cite{mybox}
that the tree contributions  dominate. Then, by taking the initial epoch
to be soon after horizon exit, one finds spectral tilt
\be
n-1 = 2\eta_{\sigma\sigma} - 2\epsilon
\dlabel{tiltfromsigma}
, \ee   
where $\eta_{\sigma\sigma}\equiv \mpl^2 (\pa^2  V/\pa\sigma_*^2) /V$

\subsection{Evolution of $\sigma$}

As noted earlier, we can usually take the relevant light fields $\phi_i(\bfx,t)$ to be smooth
on the scale $R$ even after the initial epoch. Then we can invoke the separate universe assumption,
making their evolution at each location the same as in some unperturbed universe.
This will give 
 $\sigma(\rho(t),\bfx)$ as a function of
$\rho(t)$ and $\sigma_*(\bfx)$. Just before $\zeta_\sigma$ starts to become significant
at some epoch $t\sub{start}$,
this will give some relation $\sigma(\rho(t\sub{start}),\bfx)=g(\sigma_*(\bfx))$.
  We  can  then  replace \eq{zetasigma3} by \cite{myg}
\be
\zeta_\sigma(\bfx) \simeq  \delta N(g(\sigma_*(\bfx)),\phi_*) 
. \ee
The derivatives in \eq{zetasigma2} are then evaluated using
$\pa/\pa \sigma_* = (dg/d\sigma_*)  \pa/\pa g$.

As long as $\sigma$ exchanges no energy with its surroundings, its evolution will be
given by the field equation
\be
\ddot \sigma + 3H\dot \sigma + \pa V/\pa \sigma = 0
. \dlabel{sigmafieldeq} \ee
This is a good approximation in almost all cases. (An exception within the original
curvaton scenario will occur if  $\sigma$ loses energy by preheating soon after it begins
to oscillate \cite{cnr}.)

\section{Impossibility of curvaton-type preheating with the quadratic potential}

\dlabel{schilight}


\subsection{Inflation}


\dlabel{sunpertevol}

Now we consider preheating with the quadratic potential \eqreff{pot2}.
We assume that the potential holds also during inflation, with 
the first term dominating so that $V=m^2\phi^2/2$. Under that assumption,
we are going to show that curvaton-type preheating cannot occur,
because the small value of $g$ required to make $\chi$ light during
inflation is insufficient to cause significant preheating.
In Appendix B we show that the same is true if the potential during
inflation flattens out corresponding to what is called 
inflection-point inflation.

With the potential $V=m^2\phi^2/2$, the inflaton contribution to $\zeta$ gives
$\calpzphi \simeq  m^2/\mpl^2$.
With $m^2=10^{-10}\mpl^2$ this contribution dominates, which is allowed by present
observations. We
will take the value 
$m^2 =10^{-10} \mpl^2$ as an order of magnitude estimate.
(As will become clear, curvaton-type preheating is excluded even more strongly
if $m$ is lower.)
Inflation ends when $\phi\sim \mpl$ making 
 $\phi_*^2\gg  \mpl^2$.
Therefore,  $\chi$ is light only if  $g^2 \ll m^2/\mpl^2  \sim 10^{-10}$.
We write $g_5\equiv g/10^{-5}$,  and require $g_5 \ll 1$.

During inflation the
 unperturbed fields evolve according to the slow roll approximation:
\bea
3H\dot\phi &=& -   m^2(\phi) \phi\\
 m^2(\phi) &\equiv& m^2  + g^2\chi^2  
= m^2 \( 1+ \gsn \frac{\chi^2}{\mpl^2}\) \simeq  m^2   \\
3H\dot\chi &=& - g^2\phi^2  \chi \simeq - \gsn m^2 \chi \dlabel{dotchi}
, \eea
with $3\mpl^2 H^2 \simeq m^2\phi^2/2$. 
We  assume  $\chi\ll\mpl$.
The  evolution of $\phi$ is hardly affected by $\chi$, and well before
the end of inflation $N(t) = -\frac12  \frac{\phi^2(t)}{\mpl^2}$,
where $N(t)$ is the remaining number of $e$-folds of inflation.
Using this expression one finds $\chi(t) \propto [N(t)]^{2\gsn}$,
which means that $\chi$ is practically constant during inflation. 

Inflation ends soon after the slow-roll approximation fails, which in turn
happens when $H(t)\simeq m$. 
After inflation $\phi^2$ oscillates, and  
$\chi(t) = \propto \exp[ -\gsn N(t)]$, where now $N(t)$ is the number
of $e$-folds after the end of inflation.
This behavior persists until $\phi$ decays, or the 
 mass of $\chi$ becomes significant. These events must occur
in time to create  radiation with $T\sim \MeV$, which means roughly
$N(t)\lsim 10^2$. Assuming $\gsn\lsim 10\mone$ the 
  evolution of $\chi$ is again negligible. We can therefore set $\chi(t)=\chi_*$.

At this point we mention the issue of radiative contributions to the potential,
generated by the coupling $g^2$.
According to the Coleman-Weinberg formula, the  one-loop $\xi$ contribution is
\be
V\sub{loop} = g^2 \frac{\phi^4}
{32\pi^2} \ln(\phi/Q)
, \ee
where $Q$ is the renormalization scale. To minimize higher order contributions
one should choose $Q\sim\phi_*$, and then the requirement that $V'\sub{loop}$
and $V''\sub{loop}$ are negligible corresponds to \cite{kls2}
$g^2\ll 4\pi m/\phi_* \sim 10^{-5}$.
With supersymmetry the loop correction is 
reduced and it looks as though the small
$g$ that we are invoking is safe. 
Unfortunately though, this estimate of the loop correction 
is not self-consistent because
 the renormalization scale must be below 
 the ultra-violet cutoff of the effective field theory,
which in turn must be below $\mpl$ since we are assuming Einstein gravity.
To bring  the loop correction under control one would have to 
embed the model within a larger theory. We proceed on the assumption that the
loop correction is negligible.

\subsection{Creation of $\chi$ particles}
\dlabel{subsec:chiprod}

In the literature, the potential \eqreff{pot2} is usually considered
with $g^2$ not too many orders of magnitude below 1. Then the oscillation of 
$\phi$
after inflation leads to the copious production  of $\chi$
particles. This occurs through what is called parametric resonance,
and it rapidly drains away most of the oscillation energy corresponding to
what is called preheating. We are going to show  that $g^2$ is too small for 
parametric resonance. Then we will show that 
the variation of $\phi$ around the end of inflation still creates some
$\chi$ particles, and go on to estimate the contribution of this creation
to  $\zeta_\chi$.

It is sufficient to consider  narrow resonance,  which requires the
condition  $q\equiv g^2 \phis /4m^2 \ll 1$. (The alternative regime of
broad resonance, $q\gsim 1$ requires a bigger value of $g^2$.)  Narrow
resonance occurs if  there is time for  $\phi$ to undergo some
oscillations, while the wavenumber $k/a(t)$ passes through the  narrow
band  
\be 1- q(t)/2 < \frac2 m \frac k{a(t)} < 1 + q(t)/2 . 
\ee 
The
time  for the  passage to occur is $\Delta t\simeq q/2H$, so narrow
resonance requires $mq/H  \gg 1$.  When the oscillation begins at
$H=H_*\simeq m$, this  corresponds to  $\gsn \gg 1$.  Afterwards
$\Delta t$ decreases like $a^{-3/2}$. It follows that narrow resonance
requires $\gsn \gg 1$,  which contradicts the  light field requirement
$\gsn\ll 1$. Hence {\em the light field requirement is incompatible
with parametric resonance}. The estimates of non-gaussianity from preheating 
in \cite{anupamkari1,anupamkari2} are therefore inconsinstent, because they invoke
both parameteric resonance and the assumption that $\chi$ is light.

The preheating process operates 
on scales  that never leave the horizon. 
 It occurs because
the evolution of the mode function  $\chi_k$ is non-adiabatic. At some level,
the evolution of the mode function will be non-adiabatic even if $g^2$
is too small for preheating, which means that $\chi$
particles will still be created. Let us see if this creation can give a 
significant contribution to $\zeta$.

We work with $f_k\equiv a  \chi_k$ and conformal time $\eta$. 
Denoting $d/d\eta$ by a prime, it  satisfies
 \cite{bt}
\be
f_k'' + \[ k^2 - X^2(\eta)  \] f_k = 0,\qquad
X^2 \equiv   a''/a -g^2 \phi^2 a^2 
 \dlabel{eq:f2deriv} 
\ee
To calculate the occupation number one imposes the early time  condition
$\sqrt{2k} f_k = \exp(-ik\eta)$. The late time behavior is then
\be
\sqrt{2k}f_k = \alpha_k e^{-ik\eta} + \beta_k e^{ik\eta}
, \dlabel{eq:alphabeta} 
\ee
with $|\alpha_k|^2-|\beta_k|^2=1$, and the occupation number is
$n_k = |\beta_k|^2$.

We set $\eta=0$ at the end of inflation. Well before the end of inflation,
$\eta = -1/aH\sub{end}$  and well afterwards $\eta=2/aH$. In both of these regimes,
$|\eta|\gg 1/a\sub{end}H\sub{end}$ 
 where the star denotes the end of inflation, and we
have 
$X^2\simeq a''/a\simeq 2/\eta^2$. We conclude that $X^2$ has a peak centered
on $\eta=0$ with width of order $1/a\sub{end}H\sub{end}$, and that within the peak,
$X^2\simeq a''/a \sim 1/(a\sub{end}H\sub{end})^2$. As an  approximation we may write
\be
a''/a \sim  2/ [ \eta^2 + (1/a\sub{end}H\sub{end})^2 ]
, \dlabel{peakapprox}  \ee
which becomes accurate in the regime $|\eta| \gg 1/(a\sub{end}H\sub{end})^2$.

In the regime $k\gg a\sub{end} H\sub{end}$, the square bracket in \eq{eq:f2deriv} is close to 
$k^2$
at all times, giving $n_k\ll 1$. In the opposite regime $k\ll a\sub{end} H\sub{end}$,
which we are not considering at the moment, the vacuum fluctuation
is promoted to a classical perturbation corresponding to occupation
number $n_k\gg 1$. It follows that in the regime $k\gsim a\sub{end} H\sub{end}$
under consideration, particle creation occurs mainly at the bottom end of the 
range corresponding to $k\sim a\sub{end} H\sub{end}$ and that $n_k\sim 1$ in this
range. Assuming that these particles
dominate $\rho_\chi$  we have
\be
3P_\chi=\rho_\chi \sim
\frac1{a^4} \int_0^\infty dk k^3 n_k \sim H\sub{end}^4 \( \frac{a\sub{end}}a \)^4
. \dlabel{rhochi} 
\ee
(We present in an Appendix estimates of  $n_k$ at $k\gg a\sub{end}H\sub{end}$,
showing that the contribution from this regime is indeed negligible.)
These  fall  like $a^{-4}$ since we are taking $\chi$ to be massless
and the creation occurs at $a\sim a\sub{end}$. 
The total energy density is dominated by 
\be
\rho_\phi \sim  \mpl^2 H\sub{end}^2 (a\sub{end}/a)^3
.\dlabel{rhophi} 
\ee 

\subsection{Estimating the  contribution to  $\zeta_\chi$}

The contribution $\zeta_\chi(\bfx,t)$ is given by \eq{zetasigma2}
with $\sigma=\chi$.
It is negligible just when  the smoothing scale leaves the horizon, which
we are choosing to be a bit shorter than the shortest cosmological scale.
This epoch is  50 or so $e$-folds
before the end of inflation, and we denote it by a subscript 1.

During inflation,   $\zeta_\chi(\bfx,t)$ will increase slightly, because
 the presence of $\chi$ means that there are really two slowly rolling
fields. Because we are choosing $\gsn\ll 1$ and $\chi\ll \mpl$,
this effect is negligible.
 We are  interested in the possible increase
of $\zeta_\chi(\bfx,t)$ shortly after inflation, due to the creation of
the $\chi$ particles.
We will denoting this increase by $\tilde \zeta_\chi$. It is equal to
$\delta \tilde N$, where
$\tilde N$ is  the number of $e$-folds taking place while the $\chi$
particles are being created.

Using \eq{zetadot} to first order in $\delta P$, we have
\be
\frac{\pa \tilde \zetachi}{\pa \chi_*} =
\frac{\pa \tilde N}{\pa \chi_*} = \int^{a_2}_{a\sub{end}} 
\frac{ \pa P/\pa \chi_*} {\rho + P}   \frac{da}{a}
, \dlabel{zetachi} \ee
where a subscript 2 indicates an epoch just after the end of preheating.
Without the tildes and with the integration going from the initial epoch to a
sufficiently late time, this expression gives the complete quantity 
$\pa \zeta\chi/\pa \chi_*$. It has not been given before. Note that
$\chi_*$ is evaluated at the initial epoch during the inflation as
discussed after Eq.~(\ref{eq:zeta2nd}).

To evaluate $\pa P/\pa \chi_*$, we  use 
\eqs{rhophi}{rhochi} for $\rho$ and $P$.
They depend
on $\chi_*$ because the epoch at which inflation ends depends on
the effective inflaton mass $m$, which in turn depends slightly on $\chi_*$
according to the expression $H\sub{end}^2 \simeq  m^2(\chi_*)$.
Inserting this expression into \eqs{rhophi}{rhochi}, and differentiating at
constant $\rho$,  we find
\be
\frac{\pa P}{\pa (m^2)}
= - \frac29 H\sub{end}^2 \( \frac{a\sub{end}}{a} \)^4
. \ee
Using $\pa (m^2(\chi_*)) /\pa \chi_* = g^2 \chi_*$ and \eq{zetachi} we find
\be
\frac{\pa N}{\pa \chi_*} \sim - \frac{g^2 \chi_*}{\mpl^2}
=\gsn \frac{m^2\chi_*}{\mpl^4}
. \ee

{}From these expressions we find
\be
\calpzchi/\calpzphi \sim ({\gsn})^4 
\( \frac m \mpl \)^4 \( \frac {\chi_*}{\mpl} \)^2
\ll \( \frac m \mpl \)^4 \sim 10^{-20}
. \ee
Assuming $\chi_*\gg H\sub{end}$ 
 the contribution of $(\delta\chi_*)^2$ to the non-gaussianity parameter is 
given by \eq{fnltree}
\be
\frac65f\sub{NL}^\chi = 
\( \frac{\calpzchi}{\calpzphi} \)^2 
\frac{\pa^2 N/\pa\chi_*^2}{ (\pa N/\pa \chi_*)^2 } 
\sim (\gsn)^6 \( \frac m \mpl \)^6 \( \frac {\chi_*} \mpl \)^2 \ll 
\( \frac m \mpl \)^6  \sim 10^{-30}
. \ee
If instead 
$\chi_*\lsim H\sub{end}$ the result for $f\sub{NL}^\chi$ 
is even smaller, given by 
\eq{last}. 

Of course this is not the only  contribution to $\fnl$. The contribution
$N_\chi\delta\chi_*$ will
 contribute  \cite{fieldnong} because the bispectrum of
$\delta\chi_*$ is not exactly zero \cite{slfield}.
The  dominant contribution though will be from the inflaton perturbation,
giving  \cite{mald,slfield} $|\fnl|\sim 10\mtwo$.

\section{Modulated preheating with the quadratic potential}
\dlabel{sec:sip}

To generate significant preheating with the quadratic potential
\eqreff{pot2},  we need $g^2\gg 10^{-10}$.
This makes $\chi$ heavy during inflation, and for preheating to generate a 
significant
contribution to $\zeta$ we need to introduce a third field $\sigma$. The 
contribution of
$\sigma$ to the potential should be such that (i) $\sigma$ is light during 
inflation
and (ii) its value contributes to the effect value of 
$g$ so that we have 
$g(\sigmastar)$.  Then the perturbation $\delta \sigma$
on cosmological scales will give a non-adiabatic pressure perturbation 
between the end of inflation 
and reheating, which will generate a contribution to 
$\zeta$. 
This  scenario is similar to the modulated reheating scenario
\cite{inhomreh,zaldarriaga,acker}, where the decay rate $\Gamma(\sigmastar)$ 
depends on a light field $\chi$.   For comparison, we recall
the modulated reheating result, using the $\delta N$ formalism
 \cite{newbook}.

\subsection{Modulated reheating}

For modulated reheating it is a good approximation to take
the decay as instantaneous. Then  $\zeta_\chi$ vanishes before
decay and is constant thereafter. It can easily be calculate because
the dependence of the local scale factor is known both before and
after decay \cite{inhomreh,zaldarriaga,acker,newbook}.
Before decay there is  matter domination
with  $a\propto H^{-1/2}$. 
The  decay occurs when $H=\Gamma$, therefore
\be
 \frac{a\sub{dec}}{a_{\rm i}} \propto  \Gamma^{-2/3},\qquad
\frac{a_{\rm f}}{a\sub{dec}} \propto \Gamma^{1/2}
. \ee
This gives $e^N\propto \Gamma ^{-\alpha}$ with $\alpha=1/6$. Then
\be
\frac{\pa N}{\pa \Gamma} = -\alpha \frac1\Gamma,\qquad
\frac{\pa^2 N}{\pa \Gamma^2} = \alpha \frac1{\Gamma^2}
. \ee
This gives \cite{newbook}
\be
N_\sigma = -\alpha \Gamma'/\Gamma ,\qquad
N_{\sigma\sigma} =  -\alpha \Gamma''/\Gamma + \alpha  \( \Gamma'/\Gamma \)^2
, \ee
and
\be
\calpzsigma = \left(\alpha \frac{\chi_* \Gamma'}{\Gamma}\right)^{2} 
\left( \frac
{H_*}{2\pi\sigma_*}
\right)^{2},  
\qquad
 \frac65 f\sub{NL} = - \( \frac {\calpzsigma} \calpz \)^2
\alpha \[ \frac{\Gamma'' \Gamma}{\Gamma'^2} - 1
\]. \dlabel{reheatpred} 
\ee

For illustration, one may take a simple possibility \cite{acker}:
\be
g^2(\sigmastar) = g^2 \( 1 + \frac{\sigmastar^2}{M^2} \),\qquad
g(\sigmastar) \simeq g \( 1+ \frac12 \frac{\sigmastar^2}{M^2} \) \simeq g
. \dlabel{reheatpred2} \ee
with $M$ some mass scale. Then 
\bea
\calpzchi\half &\sim& \alpha \frac{\sigmastar}{M^2} 
\( \frac{H_*}{2\pi} \) \\
\fnl &\sim& \alpha  \( \frac{\calpzsigma}{\calpz} \)^2 \( 1-
\frac{M^2}{2\chistar^2} \). 
\dlabel{preheatpred2} \eea
These expressions 
allow $\zeta_\sigma$ to dominate and they allow $\fnl$ to be  
much bigger than 1.

\begin{figure}[ht]
    \begin{center}
    \includegraphics[width=100mm,clip,keepaspectratio]
{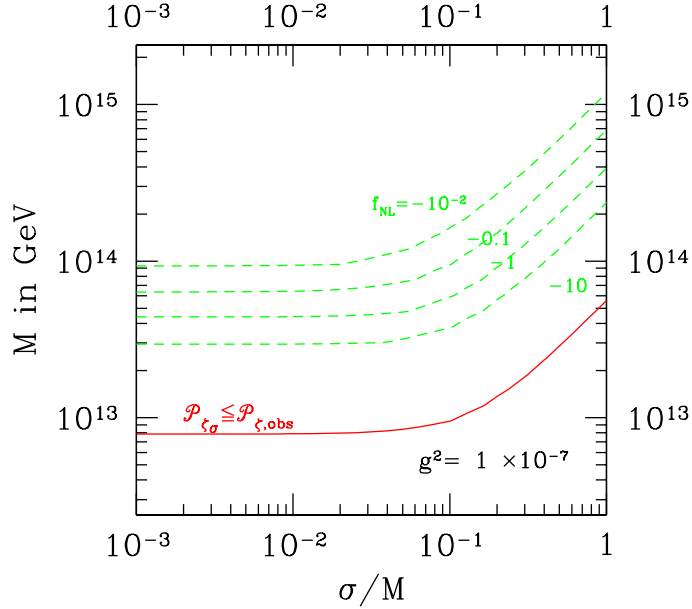}
        \caption{$f_{\rm NL}$ in the plane of $\chi/M$ and $M$
        in case of $g^{2} = 10^{-7}$. We plotted $f_{\rm NL} =
        -10^{-2}, -0.1, -1$ and $-10$ from top to bottom. The boundary
        ${\cal P}_{\zeta_{\sigma}} \le {\cal P}_{\zeta,{\rm obs}}$ (
        $\sim 3 \times 10^{-9}$) is also plotted by the solid line.}
      \dlabel{first}
    \end{center}
\end{figure}
\begin{figure}[ht]
    \begin{center}
     \includegraphics[width=100mm,clip,keepaspectratio]
 {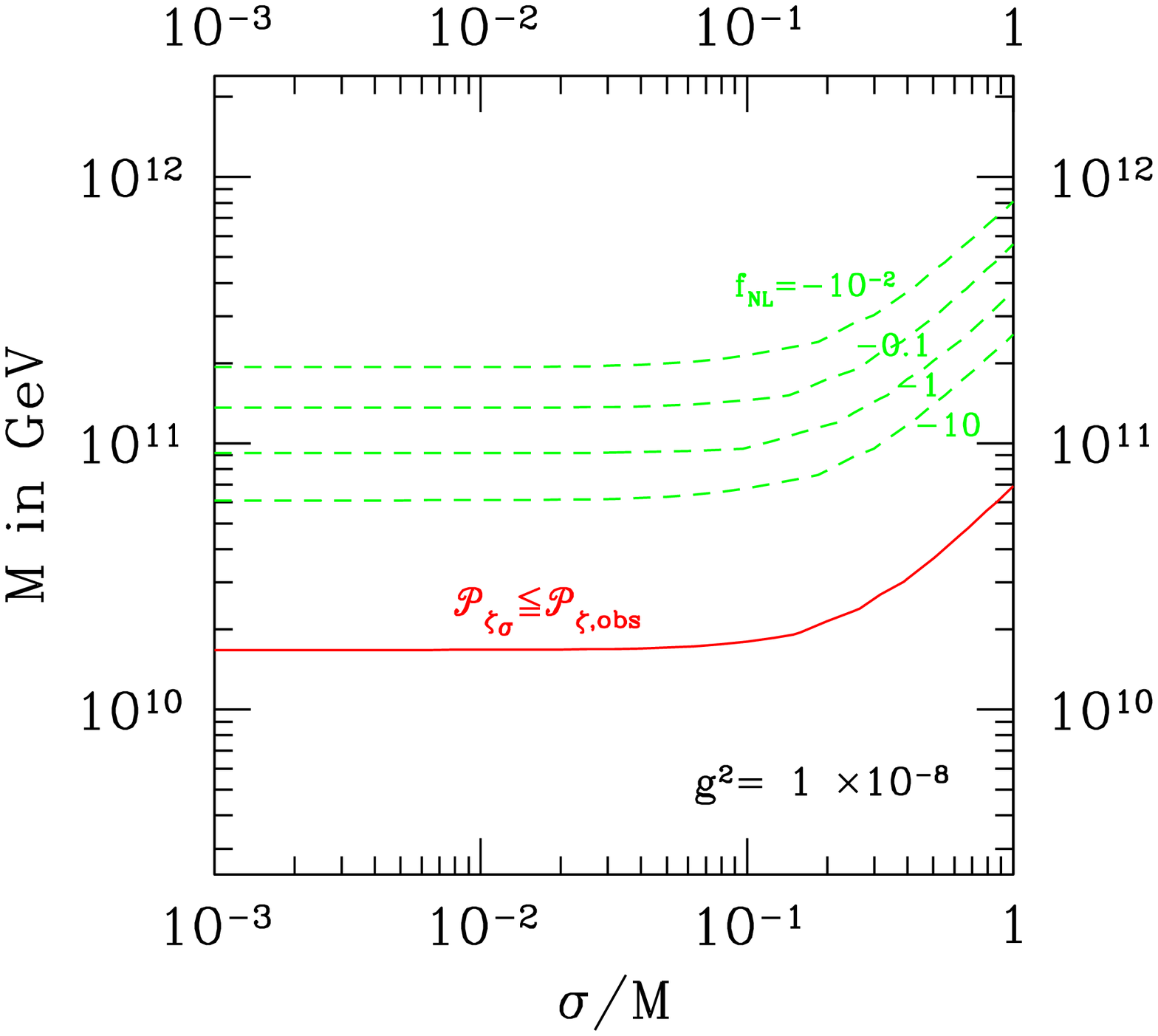}
        \caption{Same as Fig.~\ref{first} but for $g^{2} = 10^{-8}$}
      \dlabel{second}
    \end{center}
\end{figure}

\subsection{Modulated preheating}

Preheating generates a contribution to $\zeta$ which changes
continuously during preheating, and continues to change afterwards
for as long as $P_\chi$ generates a significant 
non-adiabatic pressure perturbation.  We have therefore
\be
\zeta_\sigma = \delta N = \delta N\sub{pre} + \delta N\sub{after}
, \ee
where $N\sub{pre}$ is the number of $e$-folds of preheating and 
$N\sub{after}$ is the subsequent number of $e$-folds up to an  epoch
when the non-adiabatic  pressure has become  negligible.

Given $N(\sigmastar)$ we have
\bea
\dlabel{eq;Ns}
N_\sigma  &=& g' \frac {\pa N}{\pa g}
  \\
\dlabel{eq;Nss}
N_{\sigma\sigma} &=& g'' \frac {\pa N}{\pa g}
+{g'}^2   \frac {\pa^2 N}{\pa g^2}
. 
\eea 
Inserting these formulas into \eqs{first}{last}
gives the preheating contribution to the 
spectrum and non-gaussianity of the curvature perturbation.

 For simplicity, we will assume that
$\rho_\chi$ has almost redshifted when  $\phi$ decays, so that
the non-adiabatic pressure has become negligible by that time. 
Since $\rho_\phi\propto a\mthree$ after preheating is over, this gives
\be
N\sub{after}(\bfx,t) 
= \frac13 \ln\( \frac{\rhophiend} {\rho(t)} \)
, \dlabel{nafter} \ee
where the subscript 2 indicates  
 the end of preheating and  $t$ can be taken as the 
time just
before $\phi$ decays. We then have
\be
3\frac{
\pa N\sub{after}} {\pa g} = \frac1{\rhophiend}
\frac{\pa \rhophiend}{\pa g},
\dlabel{dnafter}
 \ee 

In general,  an accurate calculation of $\rhophiend$ and $N\sub{pre}$ 
requires 
numerical simulation. Here we will just use analytic estimates
\cite{kls2}, which should be good in the regime $10^{-10} < g^2 < 10^{-7}$.
In this regime, $\rhochiend \lsim \rhophiend$, with the equality attained
for $g^2\sim 10^{-7}$. 

According to \cite{kls2}, the $\chi$ particles are non-relativistic 
during preheating.
Hence $\rho_\chi$ as well as $\rho_\phi$ is matter-dominated during this 
era
and the total energy density is proportional to $a\mthree$.
According to \cite{kls2}, preheating ends at the epoch
given by 
\be
(mt_2)^2 \sim \(\frac
{a_2}{a\sub{end}} \)^3 \sim \( \frac {g\mpl} {m} \)^2 = g_5^2
, \dlabel{aend} \ee
where $t_2 = g \sqrt{8\pi}M_{\rm P}/3m^{2}$ and $g_5 \equiv g/10^{-5}$.
Let us check that the $\chi$ particles are indeed non-relativistic up to 
this epoch.
Their  effective mass-squared is
\be
m_\chi^2(t) = g^2 \phis \sim g^2 \mpl^2  (a\sub{end}/a(t))^3 \
, \ee
where the star denotes the end of inflation. According to \cite{kls2}, 
the momentum of the $\chi$
particles in our regime is \cite{kls2} $k/a\sim \sqrt{g M_{\rm P} m} (a\sub{end}/a)$.
It follows that the $\chi$ particles are non-relativistic until
\be
(a/a\sub{end}) \sim g\frac{M_{\rm P}}m = g_{5}
, \ee
which is after the end of preheating.

{}From \eq{aend} we get
\be
\frac{\pa N\sub{pre} }{\pa g} =\frac2{3g},\qquad 
\frac{\pa^2 N\sub{pre} }{\pa g^2} = -\frac2{3g^2}. 
\dlabel{npreh}
\dlabel{eq:npreh}
\ee

To handle $N\sub{after}$ we need $\rho_\phi\su{end}(g)$.
The  total energy density at the end of 
preheating  follows from   \eq{aend}:
\be
\rho\sub{end} \sim \mpl^2 m^2 (a\sub{end}/a_2)^3 \sim m^4/g^2
. \dlabel{rhoend}\ee
Section IXA of \cite{kls2} nicely gives the expression of
$n_{\chi}(t)$ for $t < t\sub{end}$. Multiplying it by $m_\chi(t)$ to 
get $\rho_\chi$,
we find at the end of preheating
\be
\rhochiend(g) =A g\mone e^{B  g } 
,\dlabel{rhochiend} \ee
with
\bea
A&=& \frac{3^{5/2} \times 10^{-4} m^{5}}{\sqrt{8\pi} \mu^{1/2} M_{\rm P}} \\
B &=& 2\times(8\pi)\half \times 0.14\mpl /m \simeq \mpl g/m 
. \eea
The difference between \eqs{rhochiend}{rhoend}
gives $\rho_\phi\su{end}(g)$, from which we can calculate
$\pa N\sub{after}/\pa g$ and $\pa^{2} N\sub{after}/\pa g^{2}$ as follow,
\begin{eqnarray}
    \dlabel{eq:dNafterdg}
    \frac{\partial N_{\rm after}}{\partial g} 
    = \frac1{3(\rho^{\rm end} - \rho^{\rm  end}_{\chi})}
    \left( - \frac{2\rho^{\rm end}}{g} 
      - B \rho^{\rm  end}_{\chi} \right),
\end{eqnarray}
\begin{eqnarray}
    \dlabel{eq:dN2afterdg2}
    \frac{\partial^{2} N_{\rm after}}{\partial g^{2}} 
    = \frac13 
    \left[
\frac{-1}{(\rho^{\rm end} - \rho^{\rm  end}_{\chi})^{2}}
    \left( - \frac{2\rho^{\rm end}}{g} 
      - B \rho^{\rm  end}_{\chi} \right)^{2}
    + \frac1{(\rho^{\rm end} - \rho^{\rm  end}_{\chi})}
    \left( \frac{6\rho^{\rm end}}{g^{2}} 
      - B^{2} \rho^{\rm  end}_{\chi} \right)
    \right]. \nonumber \\
\end{eqnarray}
Adding these into (\ref{eq:npreh}), we obtain the total ${\pa
N}/{\pa g}$ and ${\pa^2 N}/{\pa g^2}$.

Except near the top of our range $10^{-10}< g^2 <10^{-7}$,
the ratio $R_\chi\equiv \rho_\chi\su{end}/\rho\sub{end}$ is
much less than 1. To first order in $R_\chi$ we find simply
\bea
\frac{\pa N}{\pa g} &\simeq & - (\mpl/m) R_\chi \simeq - 10^5 R_\chi  
\dlabel{ford1} \\
\frac{\pa^2 N}{\pa g^2} &\simeq & - (\mpl/m)^2  R_\chi \simeq - 10^{10}
 R_\chi 
. \dlabel{ford2} \eea 
Taking $g(\chistar)$ to be given by \eq{reheatpred2} we can calculate
$\calpzchi$ and $\fnl$ {} from \eqst{first1}{last}. In the regime where
either of them may be significant, the loop contributions dominate. 
Setting $\ln(kL)=1$ we plot the result in Figures
\ref{first} and \ref{second}. (The exact result was used instead of
\eqs{ford1}{ford2}, which is needed for Figure \ref{first}.)
 We see  that the observational bound $\fnl \gsim -10$ can be
satisfied only if  $\calpzchi/\calpz$ is  very small.  
{}

\section{Massless  preheating}

\dlabel{sctp}

In this section we consider curvaton-type preheating with the potential
\eqreff{pot3}, usually called massless preheating. 
{} Curvaton-type preheating, corresponding to a contribution $\zetachi$,
 is indeed possible for this case, because there is a  region
of parameter space which makes $\chi$ light during inflation. 
Also, the inflaton trajectory
can be in practically the $\phi$ direction. Then  we
 we deal with single-field inflation,
and  the contribution $\zetachi$ becomes significant only after 
inflation is over, ie.\ the preheating contribution to $\zetachi$ is the dominant one. 

With the potential $V=\lambda\phi^4/4$, the inflaton contribution to 
$\zeta$ gives
\cite{treview,newbook}
$\calpzphi = 10^3 \lambda$. To agree with observation it should not dominate
\cite{wmap5} which requires $\lambda \lsim 10^{-12}$. To obtain significant 
preheating one requires roughly $g^2\sim \lambda$. These small couplings
beg explanation,  and as with the quadratic potential
there is also the issue of keeping radiative contribution
to the potential under control (though the naive estimate $V\sub{loop}
\sim g^4 \phi^4 \ln(\phi/Q)$ makes that contribution  tiny).
 We proceed without addressing these issues.

\subsection{Inflation}

To keep things simple we will pretend that all cosmological scales leave 
the horizon 55 $e$-folds before the end of inflation, and denote their values
then by a subscript 55. 
Following \cite{lev} we focus on the case $g^2=2\lambda$ (corresponding to a supersymmetry).

Single field inflation occurs if $\pa V/\pa \phi\gg \pa V/\pa \chi$,
 which corresponds to $\chi\ll \phi$. We assume this condition after the biggest cosmological 
scale leaves the horizon, which occurs when $\phi\sim 10\mpl$. 
 Then,
during inflation and for some time afterwards, 
\be
\ddot \phi + 3H\dot\phi + \lambda\phi^3  = 0 
\dlabel{linphi} , \ee
with  $H^2 = \lambda \phi^4/12\mpl^2$ during inflation. 
Inflation ends at the epoch $\phi\sub{end}\simeq \mpl$.
 During inflation, $\eta_{\chi\chi} = 4\mpl^2/\phi^2$, so that $\chi$ is light 
except near the end of inflation. 

In the slow-roll approximation,  the 
 evolution of the fields is given by
\bea
-\frac{\dot \phi}{H} &=& \frac{\lambda \phi^3}{3H^2} = 
 4 \frac{\mpl^2}{\phi} \dlabel{phiev} \\
-\frac{\dot \chi}{H} &=& \frac{g^2\phi^2\chi}{3H^2} = \frac43
\frac{g^2}\lambda \frac{\mpl^2\chi}{\phi^2}  \dlabel{chiev}
. \eea
Well before the end of 
inflation, these give
\be
\phi(t) 
\simeq \sqrt{8N(t)}\phi\sub{end}\qquad \chi(t)  =
(8N(t) )^{g^2/6\lambda}\chi\sub{end} 
, \dlabel{phichioft} \ee
where $N(t)$ is the remaining number of $e$-folds of inflation. The earliest epoch
of interest is the one when the scale
$k=H_0$ leaves the horizon, corresponding to $N(t)\simeq 55$.
 Then 
\be
\phiff \simeq 21\phi\sub{end}\qquad \chiff \simeq 8 \chi\sub{end}
. \ee
 The corresponding Hubble parameter is
$\Hff\simeq 440 H\sub{end}$.
An  accurate calculation \cite{personal}, taking
account of the failure of slow roll near the end of inflation gives 
$\chiff \simeq 10^2 \chi\sub{end}$.

\subsection{Preheating in an unperturbed universe}

The  field equations are
\bea
\ddot \phi + 3H\dot\phi + \lambda\phi^3 + g^2\phi \chi^2 &=& 
a\mtwo \nabla^2 \phi \dlabel{nonlin1} \\
\ddot\chi + 3H\dot\chi + g^2\phi^2\chi &=& 
a\mtwo \nabla^2 \chi  \dlabel{nonlin2}
. \eea

In the regime $\chi_*\ll\phi_*$,  the first order perturbation of
\eq{nonlin2} gives
\be
\ddot\chi_\bfk + 3H \dot\chi_\bfk + g^2\phi^2\chi_\bfk + (k/a)^2 \chi_\bfk = 0
. \dlabel{chibfk} \ee
In the early stage of preheating, $\phi$  oscillates according to \eq{linphi}.
 This  corresponds
to a time-averaged $P=\rho/3$ (not to $P\ll \rho$ as for the quadratic 
potential).
The broad resonance band of $k$ is time-independent 
(not redshifted as for quadratic inflation)
and depends only on the combination $g^2/\lambda$. For cosmological scales to be in the band,
it has to extend down to $k=0$. That is the case for $1<g^2/\lambda <3$, which is satisfied by our
choice $g^2=2\lambda$. There is no significant preheating for $g^2 < \lambda$.

As we noted earlier, $\zeta\simeq \zetaphi$ is forbidden in this model. In the opposite case
 $\zeta_\chi\simeq \zeta$, the spectral index is given by   \eq{tiltfromsigma}.
In the slow-roll approximation \eqreff{phichioft} with $\chi=0$ this gives\footnote
{We fixed the coefficient $0.02$ by taking slow roll to be valid right up to the end of inflation,
with $\phi\sub{end}^2=8\mpl^2$ which corresponds to ending inflation at $\epsilon=1$.
A numerical calculation handling the departure from slow roll at the end of inflation
alters this estiamte by only a few percent \cite{personal}.}
\be
n-1 \simeq  0.02
 \frac{50}{N}  \( \frac{g^2}{2\lambda} - 1\)
. \ee
This too is forbidden by observation with  our chosen value $g^2=2\lambda$ but it might
agree with the minimum value $g^2=\lambda$. Otherwise we would need to assume
both $\zeta_\phi\ll \zeta$ and $\zeta_\chi\ll \zeta$, requiring a third light field to give
the dominant contribution to $\phi$.

To arrive at an estimate of $\zetachi$, we will invoke  \cite{lev}  which computes  
 the number $N\sub{pre}(\chi\sub{end},\phi\sub{end})$
of $e$-folds of preheating in a universe that is  
unperturbed at the end of inflation.
To be more precise, $N\sub{pre}(\chi\sub{end},\phi\sub{end})$
is the number of $e$-folds from the end of inflation
(when $\dot H=-H^2$) to an epoch of fixed $\rho$ soon after the end of
preheating which is around five $e$-folds later.\footnote
{At the final  epoch,  the cosmic fluid has equation of state $P/\rho=1/3$,
except for the  contribution of the
$\chi$ oscillation. The latter is small, so that $\rho(t)$ is smooth except
for a small oscillation. This small oscillation is averaged over so as to
avoid the small
oscillation in $N\sub{pre}(\chi\sub{end},\phi\sub{end})$ as a function
of $\chi\sub{end}$, that would otherwise be present. In order not to affect
the final outcome, represented by the smoothed function
$N_R(\chi\sub{end},\phi\sub{end}$ defined below, that oscillation should
have period $\ll H\sub{end}$, but we have not checked whether that is the
case.}

The calculation  is done within a comoving box, whose size
at the end of inflation is $20/H\sub{end}$, with
$H\sub{end}/2\pi=10^{-7} \times \sqrt{8\pi\mpl}$.\footnote
{Note that their  $\mpl$  is $\sqrt 8\pi$ times ours.
The value of $H\sub{end}$
is about the one that would make $\zetaphi$  equal to the observed value. 
In reality,  observation requires $\zeta_\phi$  and therefore $H\sub{end}$ to be somewhat 
smaller but that  would hardly the following analysis.}
The lowest wavenumber is then  $k=(\pi/10) (aH)\sub{end}$, which means that all $k$ come inside the 
horizon soon after  preheating begins. 

Two  methods of calculation were used,  which give similar results
\cite{lev,personal}.
 One of them uses  two stages \cite{personal}.
 The first stage, lasting for about two $e$-folds,
is done in Fourier space. The $k=0$ mode 
is evolved using the non-linear equations
\eqsref{nonlin1}{nonlin2} (with the right hand sides zero).
 The $k\neq 0$ modes are evolved analytically using \eqs{linphi}{chibfk}. 
Before each such mode leaves the horizon,
it is treated quantum mechanically so that \eq{linphi} applies to the mode function.
In this way, the vacuum fluctuation present before horizon exit becomes a classical 
perturbation after horizon exit.\footnote
{Modes with $k\ll(aH)\sub{end}$  are classical already at the end of inflation, but the small
box size practically excludes such modes, justifying our statement that the simulation starts
with an unperturbed universe.}
Up to some $k\sub{max}$ there is parameteric resonance generating
classical perturbations from the vacuum fluctuation. The second stage is a numerical simulation
of  \eqs{nonlin1}{nonlin2} using  a new code Hlattice. 
The other method does the numerical simulation from the beginning, using an 
update of the existing code DEFROST  \cite{frolov}.

 The result  using DEFROST is presented in
Figure 1.\ of \cite{lev}, with $\phi\sub{end}$ fixed at the value corresponding to $\epsilon=1$
and with $\chi\sub{end}$ in  the range
 $10^{-7}\mpl \lsim \chi\sub{end} \lsim 10^{-4}\mpl$.\footnote
{An arbitrary 
 constant is  subtracted from 
$N\sub{pre}(\chi\sub{end},\phi\sub{end})$ to make it of order $10^{-5}$, and the result is 
somewhat confusingly labelled as $\delta N$.}   
 In accordance with earlier (but not sufficiently accurate)
work \cite{kls3,massless2}, it is found that $N\sub{pre}(\chi\sub{end},\phi\sub{end})$  is not 
smooth function of $\chi\sub{end}$, but rather exhibits  sharp spikes. There are 
big spikes having a   spacing of  order $H\sub{end}$ with smaller spikes in between them.
The width of the spikes (not visible on the plot) is of order $10\mtwo H\sub{end}$ 
\cite{personal}. 

 As explained in \cite{lev}, the spikes 
 occur because the evolution of the zero mode $\{\phi(t),\chi(t)\}$ 
is given, until a fairly late stage,
by the unperturbed equations (\eqs{nonlin1}{nonlin2} with the right hands sides zero).
Each spike corresponds to a  value of $\{\phi\sub{end},\chi\sub{end}\}$
that makes $\dot\chi(t)$  anomalously small  at some epoch . 

\subsection{An estimate of  $\zeta_\chi$}

We are trying to calculate the contribution $\zeta_\chi$, of $\delta\chi_*$ to $\zeta$. It is
defined by \eq{zetasigma4}
 and we would like to arrive at the approximation \eq{zetasigma2}. 
In \cite{lev}, it is proposed that
this is done by following procedure. 
First smooth  $N\sub{pre}(\chi\sub{end},\phi\sub{end})$ (as a function of $\chi\sub{end}$
with  fixed $\phi\sub{end}$)
by a gaussian window function with variance
\be
\sigma^2_\chi \equiv (H\sub{end}/2\pi)^2 \ln[R(aH)\sub{end}]
, \dlabel{sigmadef} \ee
to obtain a function $N_R(\chi\sub{end},\phi\sub{end})$. 
(As the notation indicates, $\sigma_\chi^2$ is the 
mean square of $\delta\chi\sub{end}$,  within a region of size $R$.)   
Then write 
\be
\zeta_\chi(\bfx)  = \delta N_R(\chi\sub{end}(\bfx),\phi\sub{end})
. \dlabel{zetachix} \ee
  The smoothed function
$N_R(\chi\sub{end},\phi\sub{end})$ is  shown in Figure 1 of \cite{lev}.

We will consider the validity of this procedure in the next subsection. Adopting it for the moment,
we can use information given in \cite{lev} to estimate $\calpzchi$ and $\fnl$.
In the range
  $\chi_*\lsim 10^{-5}\mpl$, the smoothed function   $N_R$  is 
 quadratic:
\be
N_R  = f_\chi \chistar^2,\qquad f_\chi=2\times 10^6/8\pi\mpl^2
. \dlabel{quadnchi} \ee

Let us suppose first that  $\chi_*$ (the spatial average within the minimal box) vanishes.
(As we noted in Section \ref{snoninf} that is unlikely but it cannot be ruled out.)
Then \eqs{ploop}{last} give 
\bea
\calpzchi &=& 4\times 10^{-14} \( \frac{10^6 H_*}{2\pi\mpl} \)^4 \\
\fnl&=& 5\times 10\mthree \( \frac{10^6 H_*}{2\pi \mpl} \)^6
. \eea
Since $H_*/2\pi\lsim 10^{-7}\mpl$ we see that $\chi_*$ is giving a negligible
contribution. 

This calculation of $\fnl$ may be compared with the one reported in  \cite{anupamasko},
which also took the unperturbed value of $\chi_*$ to vanish.
For $\zeta_\phi$ they
 invoke the  usual  formula of first order cosmological perturbation theory,
which as is well known is manifestly the same as the $\delta N$ formula
\eqreff{calpzphi}. But for $\zeta_\chi$ they invoke second-order cosmological perturbation theory,
stopping the calculation at an epoch when  just a few  oscillations of $\phi(t)$ have taken place,
so that parametric
resonance still applies.  This should in principle give a valid result for $\zeta_\chi$
at that epoch, though one might be concerned about the complexity of the second order equations\footnote
{Analogous equations presented in a   two-field inflation model lead to a result
\cite{vaihkonen} that disagrees with that obtained from the  
$\delta N$ formalism \cite{curvatonsecondorder}. As the $\delta N$ calculation  in that case is
very straightforward, the discrepancy  in that case  presumably indicates an error in the perturbative
calculation, caused by the  complexity of the perturbative equations.}
The final step would then be to calculate  the spectrum and bispectrum of $\zeta_\chi$, to obtain
$\calpzchi$ and $\fnl$. 
That step was  not however performed in \cite{anupamasko}. Instead,
a formula $\fnl\sim \zeta_2/\zeta_1^2$ was invoked 
with $\zeta_2=\zeta_\chi$ and $\zeta_1=\zeta_\phi$.   But this formula applies  only if
$\zeta$ is of the local form, ie.\ if  $\zeta_2(\bfx)= (3/5)\fnl \zeta_1^2(\bfx)$. In the present
case where $\zeta_1$ and $\zeta_2$ are uncorrelated  it obviously cannot apply. Neither does
it  apply  if the quantities
are replaced by typical magnitudes so that it becomes $\fnl\sim \calpzchi\half/\calpzphi$.
The correct estimate   \cite{bl} is in fact $\fnl\sim \calpzchi\threehalf/\calpzphi^2$,
corresponding to \eq{last}.

So far we took $\chi_*$ to vanish. Now assume instead a more likely value
 $\chi_*\gg  H_*$.  Then \eqs{treespec}{fnltree} apply, giving  
\bea
\calpzchi &=& 4\times 10^{-12} \( \frac{10^5 \chi_*}\mpl \)^2
\( \frac{10^6H_*}{2\pi} \)^2 \\
\fnl &=& 5\times 10\mone
 \( \frac{10^5 \chi_*}\mpl \)^2
\( \frac{10^6H_*}{2\pi} \)^4 
. \eea
According to Fig.~1 of \cite{lev}, 
the quadratic approximation holds only out to $\chistar\sim 10^{-5}\mpl$, which means
that $\calpzchi$ and $\fnl$ are again negligible.
We conclude  that if the  unperturbed $\chistar$ is
small enough for \eq{quadnchi} to apply, the curvature perturbation 
in this scenario is still negligible at the end of preheating.

In the range  $10^{-5}\lsim \chi_*/\mpl \lsim 10^{-4}$, the
 $N_R$ presented in Figure 1 of \cite{lev} is an oscillating function of $\chi\sub{end}$.
As one would expect, the oscillation is slow so that $N_R$ is still a smooth function
of $\chi\sub{end}$ over an interval $\Delta\chi\sub{end} \sim H\sub{end}$.
Therefore, the usual approximation 
 \eq{zetachi} should still be  good.
The corresponding spectrum   and bispectrum have not yet been
calculated, and the calculation of $N(\chistar)$
has not yet been extended to larger $\chistar$, but the smoothness of $N_R$ and
hence the validity of  \eq{zetachi} should still hold.
Observational consequences of its failure are mentioned in
in \cite{lev}.

\subsection{Justifying the estimate of $\zetachi$}

To formulate an approximation within which \eq{zetachix}  can be   justified we proceed in two
steps. First we formulate an approximation within which $N\sub{pre}(\chi\sub{end}(\bfx),\phi\sub{end})$
can be regarded as the number of $e$-folds of preheating at a given location.
Then we justify the smoothing procedure that gives $N_R(\chi\sub{end}(\bfx),\phi\sub{end})$.

The first step requires the existence of a scale $L\sub{hom}$, which is big enough to be well outside
the horizon at the end of preheating, yet small enough that the variation of
$ N\sub{pre}(\chi\sub{end}(\bfx),\phi\sub{end})$ is negligible over a distance $L\sub{hom}$.
Let us consider these  criteria in turn.

Taking $a\propto t\half$ during preheating (corresponding to roughly
$\rho\simeq P/3$) and assuming $N\sub{pre}\sim 5$
 $e$-folds of preheating, the first requirement becomes
\be
L\sub{hom}\gg e^{N\sub{pre}} H\sub{end}\mone
. \label{firstreq} \ee
For the second requirement, we can 
 take `negligible' to mean `much less than the height $\Delta N\sub{peak}$
of a typical peak
in $N\sub{pre}(\chi\sub{end})$. The variation of $N\sub{pre}(\chi\sub{end}(\bfx))$ within a distance
$L\sub{hom}$ is 
\be
\Delta N \sim \frac{dN\sub{pre} }{d\chi\sub{end}} |\del \chi\sub{end} | L\sub{hom}
. \ee 
The maximum of $dN\sub{pre}/d\chi\sub{end}$ is achieved near a peak, and is of order $\Delta N\sub{peak}
/\Delta \chi\sub{peak}$ where $\Delta \chi\sub{peak}$ is the width of the peak.
The requirement $\Delta N\ll \Delta N\sub{peak}$ is therefore
\be
|\del \chi\sub{end}| L\sub{hom} \ll \Delta \chi\sub{peak}
. \ee
Using \eq{msgradient} to estimate $|\del\chi\sub{end}|$, this requires $\chi\sub{end}$ to be smooth on a 
scale $R$ satisfying
\be
R  \gg H\sub{end} L\sub{hom} / \Delta \chi\sub{peak}
. \ee
Using \eq{firstreq} this becomes
\be
R (aH)\sub{end}  \gg \frac{H\sub{end}} { \Delta \chi\sub{peak} }  e^{N\sub{pre}}
. \ee
Using the estimate 
\cite{personal}  $\Delta\chi\sub{peak} \sim 10\mtwo H\sub{end}$, with $N\sub{pre}\simeq 5$,
the right hand side is of order $e^{10}$ which means that $\chi\sub{end}$ should be smooth
a  scale of order $e^{10}(aH)\mone\sub{end}$. 
To achieve this,  we have to drop
the perturbations of $\chi\sub{end}$ that are generated during the last 10 or so 
$e$-folds of inflation.  There seems to be no justification for this, but let us anyway
move on to the second step.

We now have an approximation which makes  $N\sub{pre}(\bfx)$
the number of $e$-folds of preheating at a given location. But $N\sub{pre}(\bfx)$ is 
s not yet the quantity   whose  perturbation is $\zeta_\chi$.
The reason is that the latter is, by definition, smooth on the chosen scale $R$ whereas
$N\sub{pre}(\bfx)$ is not because of the narrow spikes. 
Before using the formula $\zeta_\chi=\delta N\sub{pre}$, we must 
 smooth $N\sub{pre}(\bfx)$
 on the scale $R$. We are now going to argue that the result of this smoothing will 
be approximately $N_R(\bfx)$.

Smoothing on the scale $R$ at a given location $\bfx$ means that the quantity is replaced by its
average within a sphere of radius $R$. At a random location within such a sphere, 
$\chi\sub{end}$ has a gaussian probability distribution with mean square $\sigma_\chi^2$
given by \eq{zetachix}. If $\chi\sub{end}(\bfx)$ had negligible correlation length (one much less
than $R$) the distribution at different locations would be independent and we would immediately
obtain the desired result. Unfortunately, the correlation length is practically infinite
because the spectrum of $\chi\sub{end}(\bfx)$ 
is practically flat at long wavelengths. We therefore have to proceed
differently.

We need some notation. For any function $g$
of a  variable $y$, let us write the smoothed quantity as
\be
g(y,\sigma_y) \equiv \int^\infty_{-\infty} dy' g(y') W(y'-y,\sigma_y)
, \label{smoothdef} \ee
where the window function satisfies\footnote
{ Equivalently, one can replace the third condition by $W\simeq 1$, and divide 
\eq{smoothdef} by the left hand side of the first condition.}
\bea
\int^\infty_{-\infty} dy' W(y'-y,\sigma_y) &=& 1 \\
W&\simeq& 0 \qquad (|y'-y|\gg \sigma_y ) \\
W&\simeq& \mbox{const}  \qquad (|y'-y|\ll \sigma_y )
. \eea
The window function is usually taken to be either a theta function (top hat) or a gaussian.
In any case, $\sigma_y$ is usually taken to be a constant. Then the convolution theorem
shows that smoothing kills the Fourier components of $g$ for wavenumbers $k\gg \sigma_y$
while hardly affecting them for $k\ll\sigma_y$. The same will be true even if $\sigma_y$
depends on $y$ and/or  $y'$ provided that its variation is negligible in the regime
$|y'-y|\ll \sigma_y$.

Let us denote $N(\chi\sub{end},\phi\sub{end})$ simply by
$N(\chi)$, and $N_R(\chi\sub{end},\phi\sub{end})$ by $N_R(\chi)$. We have
\be
N_R(\chi) \equiv \int d\chi' N(\chi') W(\chi'-\chi,\sigma_\chi)
, \ee
with  $\sigma_\chi$ given by \eq{sigmadef}. We  
 want to show that $N_R(\chi(\bfx))$ is smooth on the scale $R$.
Let us first assume that $\del \chi$
is practically constant within a region of size $R$. 
Then, choosing the $z$ axis in the direction of $\del \chi$, 
\be
\chi(\bfx') - \chi(\bfx) \simeq \frac{\pa \chi(\bfx)}{\pa z} 
\( z' - z \)
, \ee
where $\bfx'$ has coordinates $(x,y,z')$. We then have
\be
N_R(\chi(\bfx)) \simeq \int dz' N(\chi(\bfx')) W_R(z'-z,\sigma_z)
, \ee
where
\bea
W_R(z'-z,\sigma_z) &\equiv& W(z'-z,\sigma_z) (\pa\chi(\bfx)/\pa z)\mone \\
\sigma_z(\bfx)  &\equiv & \sigma_\chi (\pa \chi(\bfx)/\pa z)\mone
. \eea  
{}From \eq{msgradient} we learn that $\sigma_z(\bfx)\simeq R$ at a typical position.
Therefore, as required, $N_R(\chi(\bfx))$ is just the original function $N(\chi(\bfx))$
smoothed  on the  scale $R$.

We have assumed that $\del \chi$ is practically constant within a typical region of size $R$.
To see whether this is reasonable, we can estimate the mean squares of the second
partial derivatives $\pa_i\pa_i \chi$, whose Fourier components are $-k_i k_j$
times those of $\chi$. Analogously with the calculation leading to \eq{msgradient}, we find
that the typical fractional variation of $\del \chi$ within the region is of order 1.
To handle this order 1 variation, we could use curvilinear coordinates with $z$ 
constant on each spatial slice of constant $\del\chi$, and our  conclusion about 
the Fourier components of $N_R(\chi(\bfx))$ would still hold.
We have not considered the case where $\del\chi$ has very large variation, since it will
apply only to rare locations that should not affect our  conclusion.

The main source of error in this prescription for $\zeta_\chi(\bfx)$ is the neglect of the
contribution to 
$\delta\chi\sub{end}$, that is  generated from the vacuum fluctuation after $R$ leaves
the horizon.
We have no idea how to estimate this error, except by performing
a numerical simulation in a box with size rather bigger than $e^5(2\pi/(aH)\sub{end})$
that would support those modes. Even with such a simulation in place, there is at least in
principle the  fundamental  problem, that  the result for $\zeta_\chi(\bfx)$ would depend
on the chosen realization of the small-scale perturbations. One may hope that this dependence 
is small. If it is not, the derivative $N_\chi$ would become a stochastic quantity, 
and instead of $\calpzchi=N_\chi^2 \calp_\chi$ one would have
$\calpzchi\su{tree} = \vev{N_\chi^2} \calp_\chi$. Similarly, in calculating say 
$\fnl\su{tree}$ one would have to replace $N_{\chi\chi}/N_\chi^2$ by its expectation value.
These expectations values would have to be calculated using the numerical simulation.

\section{Conclusion}

\dlabel{sconclusion}

In this paper we have carefully addressed some issues, that arise when
one uses the $\delta N$ formalism is used for preheating.  Then we saw how
things worked out in a couple of specific cases. 

The investigation seems worth continuing, in a number of directions.
Several of the  the curvaton-type preheating
scenarios listed in the Introduction have been explored only with 
cosmological
perturbation theory. It seems worthwhile to look at them also with the 
$\delta N$ formalism, especially in cases where second order perturbation
theory has been used. The other paradigm, modulated preheating, would
also be worth exploring, starting with a numerical simulation for the
simplest setup using the quadratic potential of \eq{pot2}.

All of this assumes that the curvature perturbation is generated exclusively
by one or more scalar field perturbations. As has recently been realised
\cite{vec1,vec2,vec3,vec4}, 
it   might instead be generated, wholly or partly, by
one or more vector field perturbations. A smoking gun for such a setup would
be statistical anisotropy \cite{vec2,vec3,vec4}. The vector field  
possibility has 
so far been explored only with the generation of $\zeta$ taking place
through the curvaton mechanism before a second reheating \cite{vec1,vec3},
during inflation \cite{vec3} or  at the end of inflation \cite{vec2}. 
It is clear that one could implement modulated preheating using
a  vector field perturbation, allowing one or more of the parameters
to depend on the vector field in the spirit of \cite{vec2}.
One  might  also implement curvaton-type preheating using such a 
field, provided that the preheating can 
create a vector as opposed to a scalar field. It is not known at present 
whether that is the case.

\section{Acknowledgments}
During the main part of this work,  D.H.L. and K.K were   supported by PPARC grant PP/D000394/1.
D.H.L. is supported by
 EU grants MRTN-CT-2004-503369 and MRTN-CT-2006-035863.
C.A.V. is supported by COLCIENCIAS grant
No. 1102-333-18674 CT-174-2006 and   DIEF (UIS) grant No. 5134.
D.H.L acknowledges valuable discussions and
correspondence with Zhiqi Huang, which is reflected in Sections IIB and VI.


\appendix

\section{Occupation number of  $\chi$ quanta}

In this Appendix,  we discuss the solution of Eq.~(\ref{eq:f2deriv}),
\begin{eqnarray}
  f_k'' + k^2  f_k = X f_k,\qquad  X \equiv -g^2 \phi^2 a^2 + a''/a
, \dlabel{eq:f2deriv2}
\end{eqnarray}
and estimate the occupation number $n_{k}$ of the $\chi$ quanta.  
We work 
in the regime $k\gg a_* H_*$ where the star denotes the end of inflation.
According to the estimates after \eq{eq:alphabeta}, 
 $X f_k$ is a small perturbation in this regime,
which we treat to first order. We write
\be
    \dlabel{eq:defdeltaf}
    f_{k} = f_{k,0} + \delta f_k,
\ee
where $f_{k,0} = \exp(-ik\eta)/\sqrt{2k}$ is the unperturbed quantity.
When this equation is substituted into Eq.~(\ref{eq:f2deriv2}), we get a
following equation at the first order of the perturbation,
\begin{eqnarray}
    \dlabel{eq:deltafkeq}
    \delta f_{k}'' + k^{2} \delta f_{k} = X_{k}(\eta),\qquad
X_k \equiv X f_{k,0}
\end{eqnarray}

To solve Eq.~(\ref{eq:deltafkeq}), we use the Fourier transformation
method which requests  variables to be expanded into their Fourier
modes,
\begin{eqnarray}
    \dlabel{eq:fouriedeltaf}
    \delta f_{k} (\eta) = \frac1{2\pi} 
    \int^{\infty}_{-\infty} d  \omega~
    \tilde{\delta f_{k}} (\omega)~e^{i \omega \eta},
\end{eqnarray}
and
\begin{eqnarray}
    \dlabel{eq:fouriedeltaX}
    X_{k} (\eta) = \frac1{2\pi} 
    \int^{\infty}_{-\infty} d  \omega~
    \tilde{X_{k}} (\omega)~e^{i \omega \eta}.
\end{eqnarray}
In reverse, we can define the inverse-Fourier transformation of $X_{k}$
by
\begin{eqnarray}
    \dlabel{eq:fouriedeltaXtilde}
    \tilde{X_{k}} (\omega) =
    \int^{\infty}_{-\infty} d  \eta'~
    X_{k}(\eta')~e^{-i \omega \eta'}.
\end{eqnarray}
Substituting Eqs.~(\ref{eq:fouriedeltaf})~and~(\ref{eq:fouriedeltaX})
into Eq.~(\ref{eq:deltafkeq}), we immediately get a relation between
the Fourier modes,
\begin{eqnarray}
    \dlabel{eq:reltildes}
    \tilde{\delta f_{k}} (\omega) =
    \frac{\tilde{X_{k}} (\omega)~}{-\omega^{2} + k^{2}}.
\end{eqnarray}
Therefore we finally have 
\begin{eqnarray}
    \dlabel{eq:deltafkfinal}
    \delta f_{k} (\eta) = \frac1{2\pi} 
    \int^{\infty}_{-\infty} d  \omega~
    ~\frac{\tilde{X_{k}} (\omega)~}{-\omega^{2} + k^{2}} e^{i \omega \eta}.
\end{eqnarray}

\begin{figure}[ht]
    \begin{center}
    \includegraphics[width=110mm,clip,keepaspectratio]{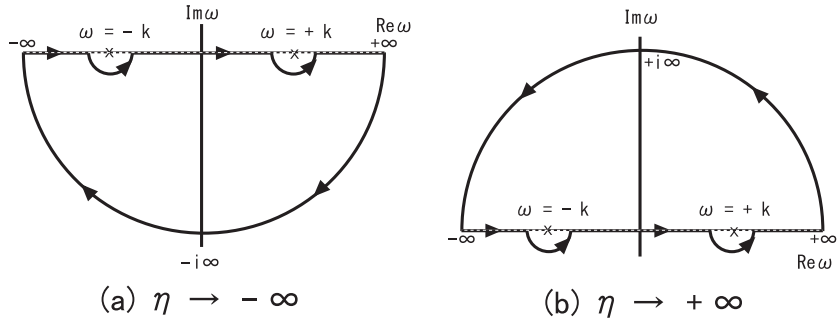}
        \caption{Integral half-circle and the position of the poles 
        in the complex $\omega$ plane. The direction of the integral
        is denoted by arrows}
      \dlabel{fig:pole}
    \end{center}
\end{figure}

The integrand has two poles, and to impose the   condition $\delta
f_k=0$ at $\eta\to -\infty$  we deform the path $-\infty < \omega
<\infty$ so that the poles lie below the path as shown in
Fig.~\ref{fig:pole} (a).  At $\eta\to -\infty$ we obtain the required
initial condition by closing the contour in the lower half plane. To
obtain instead the behavior at $\eta\to+\infty$ we close it instead
in the upper half plane shown in Fig.~\ref{fig:pole} (b) which gives
the behavior \eq{eq:alphabeta} with
\begin{eqnarray}
    \dlabel{eq:betak}
    \beta_{k} = 
    \frac{i}{(-2k)} 
    \int_{-\infty}^{\infty} d \eta
    e^{-2i k \eta} X(\eta)
\left(- g^{2} \phi^{2} a^{2}  + \frac{a''}a \right).
\end{eqnarray}

Using the approximation \eqreff{peakapprox} we find a standard Fourier
transform, which gives
\be |\beta_k|^2  = \pi^2  (a_* H_*/k)^2  e^{-4 |k|/a_* H_*},\qquad
(k\gg a_*H_*). 
\ee 
Putting this into \eq{rhochi} we find that it gives a negligible
contribution to $\rho_\chi$, 
as  advertised in the text. One might be concerned that 
the exact result for  $|\beta_k|^2$ might fall off more slowly at large
$k$ giving a significant or divergent contribution to $|\beta_k|^2$,
but the following argument should allay such concern. Since $X$ is infinitely
differentiable,  integration by parts shows that
\be
\beta_k =\frac i{-2k(2k)^n} \int^\infty_{-\infty} d\eta e^{-2ik\eta} 
\frac{d^n X}{d\eta^n}
, \ee
for all $n\geq 0$. For low $n$,  it is reasonable to extend 
the argument leading to  \eq{peakapprox}, to arrive at an estimate
\be
X^{(n)} \sim \frac1 { (a_*H_*)^n } \frac 1 { \eta^2 + (1/a_*H_*)^2 }
. \ee
This gives
\be
|\beta_k|^2 \ll  (a_* H_*/k)^{2+n}, \qquad (k\gg a_*H_*)
. \ee
Using this as  a reasonable approximation for $n=3$, we again
find a negligible contribution to $\rho_\chi$.

\section{Preheating after inflection point inflation}
\dlabel{sec:A-term}

In the main text we have studied the preheating only in chaotic
inflation models, where the potential $V=\frac12 m^2\phi^2$ generating the 
oscillation holds also during inflation. In that case
$\phi_*$ (the value at the beginning of the oscillation, or order its value
at the end of inflation)
and $m$ are given by $\phi_*\sim \mpl$ and $m^2\lsim 10^{-10}$, with the latter
inequality saturated if $\zetaphi$ 
is to be a significant fraction of the total.
Also, the Hubble parameter at the beginning of the oscillation, given
by  
\be
3H_*^2 = \frac12 \phi_*^2 m^2/  \mpl^2
\dlabel{hstarofm}
\ee
is of order $m$.

In a different class of models, known as inflection point models, the 
potential flattens out at $\phi >\phi_*$ so that its first and second derivatives
are close to zero during inflation. Then $\phi_*$ and $m$ are independent 
parameters and $H_*$ given by \eq{hstarofm} is much smaller than $m$.
We now analyze preheating in these models, assuming the interaction
$\frac12 g^2\phi^2 \chi^2$. Of course the viability of that interaction
and the possible identity of $\chi$
should be checked within a particular setup.

Here we see what happens if  $\phi_0$ is  smaller. The discussion may be
relevant for of inflation,  either in the context
of MSSM inflation \cite{mssm} which have $\phi_0\sim \TeV$ or in the context of
colliding brane inflation \cite{coll} where $\phi_0$ might be anywhere in the range
$\TeV \lsim \phi_0\lsim \mpl$. Of course, one has to check within a specific
model if the  interaction term $g^2\phi^2\chi^2$ is consistent and reasonable.

The  Hubble constant at the end of inflation is 
\begin{eqnarray}
    \dlabel{eq:HAterm}
    H_* \sim  \frac{m_{\phi} \phi_0} { M_{\rm P}}.
\end{eqnarray}

When we consider the preheating which would be induced by an
interaction term such as the second term in
Eq.~(\ref{pot2})~\footnote{Effectively we wrote the interaction term
like this for simplicity}, the effective mass of $\chi$ is given by
$m_{\chi}^2 =  g^{2} \phi_0^2$. We can parameterize the amplitude of the
oscillation term in the Mathieu equation as~\cite{kls}
\begin{eqnarray}
    \dlabel{eq:qvalueAterm}
    q = g^2  \frac{\phi_0^2}{m_\phi^2}.
\end{eqnarray}

First of all, we shall consider a condition for lightness of $\chi$
during inflation, $m_{\chi}^2  \ll  H_*^2$. Thus, it is found that
\begin{eqnarray}
    \dlabel{eq:g2upper}
    g^2   \phi_0^2  \ll  \frac{m_{\phi}^2 \phi_0^2}{ M_{\rm P}^2},
\end{eqnarray}
which means 
\begin{eqnarray}
    \dlabel{eq:eq:g2upper2}
     g^2 &\ll&   \frac{m_{\phi}^2} {M_{\rm P}^2}  \sim   10^{-30}
     \left( \frac{m_{\phi}}{\rm TeV}\right)^{2}.
\end{eqnarray}
Here only in case of MSSM inflation, $m_{\phi} \sim$ TeV.

Next we discuss the condition for successful parametric resonance. As
was discussed in Section~\ref{subsec:chiprod}, the parametric
resonance occurs if $k/a$ can stay in the  resonance band  for a
sufficiently-long time to oscillate many times.  The time interval
when $k/a$ passes the resonance band is given by $\Delta t \sim q /
H_*$. Thus that condition is represented by $z  = m_{\phi}  \Delta t
\gg 1$, which gives
\be
    \dlabel{eq:zgg1}
    q  \frac{m_{\phi}}{ H_*} = g^2 \frac{ \phi_0 M_{\rm P}}{ m_{\phi}^2} 
               \gg 1
. \ee
Then we have
\begin{eqnarray}
    \dlabel{eq:g2lower}
    g^2  \gg  \frac{m_{\phi}^2}{ \phi_0 M_{\rm P}}.
\end{eqnarray}
This lower bound on $g^{2}$ is much bigger than the upper bound
required by the lightness given in Eq.~(\ref{eq:eq:g2upper2}), when
$\phi_{0} \ll M_{\rm P}$ as is the case for the A-term inflation.


\end{document}